%% file: main.tex
\documentclass[runningheads,envcountsame]{llncs}

\makeatletter
\RequirePackage[bookmarks,unicode,colorlinks=true]{hyperref}%
   \def\@citecolor{blue}%
   \def\@urlcolor{blue}%
   \def\@linkcolor{blue}%

\def\orcidID#1{\href{http://orcid.org/#1}{\protect\raisebox{-1.25pt}{\protect\includegraphics{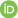}}}}
\makeatother

\usepackage{xcolor,colortbl}
\usepackage{mathtools}
\usepackage{multirow}
\usepackage{stmaryrd}
\usepackage{cancel}
\usepackage{amsmath,amssymb}
\usepackage{thmtools, thm-restate}
\usepackage[shortlabels]{enumitem}
\usepackage{microtype}
\usepackage{wrapfig}
\usepackage{pbox}
\usepackage{marvosym}
\usepackage{listings}
\usepackage[ruled, vlined, linesnumbered]{algorithm2e}
\hypersetup{
	pdftitle={A Dependency Pair Framework for Relative Termination of Term Rewriting}, colorlinks=true, linkcolor=blue, citecolor=olive, filecolor=magenta, urlcolor=cyan
}
\usepackage{todonotes}
\usepackage{placeins}
\usepackage{float}
\usepackage{tikz}
\usepackage{caption}
\usepackage{subcaption}
\usepackage{mymatrix}
\usepackage{nicefrac,xfrac}
\usetikzlibrary{shapes,calc,arrows,automata,decorations.pathmorphing,backgrounds}
\usepackage[all,cmtip]{xy}

\RequirePackage{makecell}

\usepackage[capitalize,nameinlink]{cleveref}
\usepackage{IEEEtrantools}

\newenvironment{myproof}{
	\noindent{\it Proof.}
}{\qed
	\medskip
}
\input{commands}

\SetKwInOut{Input}{input}
\SetKwInOut{Output}{output}

\definecolor{Gray}{gray}{0.85}
\definecolor{LightCyan}{rgb}{0.88,1,1}

\newcolumntype{a}{>{\columncolor{Gray}}c}
\newcolumntype{b}{>{\columncolor{white}}c}

 \newcommand{\makeproof}[2]{}
 \newcommand{\paper}[1]{}
 \newcommand{\report}[1]{#1}
 \newcommand{\final}[1]{}

 \report{
  \setlength{\textwidth}{127mm}
   \setlength{\textheight}{198mm}
 }

 \newcommand{\notnew}[1]{}
 
\paper{\usepackage[ hyperref=true, backend=bibtex, firstinits=true, maxbibnames=99, sortcites, style=numeric-comp ]{biblatex}
\addbibresource{biblio.bib}}

\report{\usepackage{cite}}

\title{A Dependency Pair Framework for Relative Termination of Term Rewriting\thanks{funded by the
    Deutsche Forschungsgemeinschaft (DFG, German Research Foundation) - 235950644 (Project
    GI 274/6-2) and DFG Research Training Group 2236 UnRAVeL}}
\titlerunning{Dependency Pairs for Relative Termination}
\author{
    Jan-Christoph Kassing\final{$^{(\mbox{\Letter})}$}\orcidID{0009-0001-9972-2470}
    \and Grigory Vartanyan\final{$^{(\mbox{\Letter})}$}\orcidID{0009-0009-9631-8307}
    \and Jürgen Giesl\final{$^{(\mbox{\Letter})}$}\orcidID{0000-0003-0283-8520}}
\institute{LuFG Informatik 2, RWTH Aachen University, Aachen, Germany}
\authorrunning{J.-C.\ Kassing, G.\ Vartanyan, J.\ Giesl}

\begin{document}
\allowdisplaybreaks

\maketitle

\vspace*{-.1cm}

\input{abstract}
\input{introduction}
\input{reltrs}

\input{dp}
\input{reldp}
\input{ADPframework}
\input{evaluation}

\paper{\printbibliography}
\report{\bibliographystyle{splncs04}
 \input{main.bbl}

}

\report{
  \clearpage

\appendix
\input{appendix}
}
\end{document}

%% file: commands.tex
\def\namedlabel#1#2{\begingroup
    #2%
    \def\@currentlabel{#2}%
    \phantomsection\label{#1}\endgroup
}

\newcommand{\aprove}{\textsf{AProVE}}
\newcommand{\muterm}{\textsf{MU-TERM}}
\newcommand{\natt}{\textsf{NaTT}}
\newcommand{\ttttwo}{\textsf{T\kern-0.15em\raisebox{-0.55ex}T\kern-0.15emT\kern-0.15em\raisebox{-0.55ex}2}}

\newcommand{\mnm}{\textsf{MultumNonMulta}}
\newcommand{\matchbox}{\textsf{Matchbox}}

\renewcommand{\emph}[1]{\index{#1}\textit{#1}}

\renewcommand{\emptyset}{\varnothing}

\newcommand{\IN}{\mathbb{N}}

\newcommand{\C}[1]{\mathcal{#1}}

\makeatletter
\def\moverlay{\mathpalette\mov@rlay}
\def\mov@rlay#1#2{\leavevmode\vtop{%
   \baselineskip\z@skip \lineskiplimit-\maxdimen
   \ialign{\hfil$\m@th#1##$\hfil\cr#2\crcr}}}
\newcommand{\charfusion}[3][\mathord]{
    #1{\ifx#1\mathop\vphantom{#2}\fi
        \mathpalette\mov@rlay{#2\cr#3}
      }
    \ifx#1\mathop\expandafter\displaylimits\fi}
\makeatother

\newcommand{\Proc}{\operatorname{Proc}}

\newcommand{\TSet}[2]{\mathcal{T}\left(#1,#2\right)}

\newcommand{\VSet}{\mathcal{V}}

\renewcommand{\O}{\mathcal{O}}
\newcommand{\R}{\mathcal{R}}

\newcommand{\DPair}[1]{\mathcal{DP}(#1)}

\newcommand{\DP}[1]{\mathtt{DP}(#1)}

\newcommand{\ADPair}[2]{\mathcal{A}_{#1}(#2)}
\newcommand{\ADPairMain}[1]{\mathcal{A}_{1}(#1)}
\newcommand{\ADPairBase}[1]{\mathcal{A}_{2}(#1)}

\newcommand{\rootsym}{\operatorname{root}}

\newcommand{\Pol}{\operatorname{Pol}}

\newcommand{\Com}[1]{\mathsf{c}_{#1}}

\newcommand{\PP}{\mathcal{P}}
\newcommand{\QQ}{\mathcal{Q}}
\newcommand{\SSS}{\mathcal{S}}

\renewcommand{\ts}{\mathsf{s}}

\renewcommand{\O}{\mathcal{O}}
\newcommand{\tf}{\mathsf{f}}
\newcommand{\tg}{\mathsf{g}}

\newcommand{\ta}{\mathsf{a}}
\newcommand{\tb}{\mathsf{b}}
\newcommand{\tc}{\mathsf{d}}
\newcommand{\td}{\mathsf{d}}

\newcommand{\tm}{\mathsf{m}}

\newcommand{\tminus}{\mathsf{minus}}
\newcommand{\tdiv}{\mathsf{div}}
\newcommand{\tdivl}{\mathsf{divL}}

\newcommand{\tcons}{\mathsf{cons}}

\newcommand{\tnil}{\mathsf{nil}}

\newcommand{\tset}{\mathsf{mset}}

\newcommand{\tM}{\mathsf{M}}
\newcommand{\tD}{\mathsf{D}}
\newcommand{\tDL}{\mathsf{DL}}
\newcommand{\tF}{\mathsf{F}}

\newcommand{\tA}{\mathsf{A}}

\newcommand{\xs}{\mathit{xs}}

\newcommand{\zs}{\mathit{zs}}

\crefname{definition}{Def.}{Def.}
\crefname{example}{Ex.}{Ex.}
\crefname{counterexample}{Counterex.}{Counterex.}
\crefname{appendix}{App.}{App.}
\crefname{ex}{Ex.}{Ex.}
\crefname{theorem}{Thm.}{Thm.}
\crefname{lemma}{Lemma}{Lemmas}
\crefname{remark}{Rem.}{Rem.}
\crefname{section}{Sect.}{Sect.}
\crefname{subsection}{Sect.}{Sect.}
\crefname{subsubsection}{Sect.}{Sect.}
\crefname{line}{Line}{Lines}
\crefname{corollary}{Cor.}{Cor.}
\crefname{figure}{Fig.}{Fig.}
\crefname{enumi}{}{}
\crefname{algorithm}{Alg.}{Alg.}

\makeatletter
\NewDocumentCommand{\dparrow}{+O{} +O{0.5cm}}{%
\begin{tikzpicture}[baseline=-0.5ex] {
\node[inner sep=0](@1) at (-0,0) {};
\node[inner sep=0](@2) at (#2,0) {};
\draw [arrows={-Triangle[open]},shorten >= 1pt,shorten <= 1pt](@1)--(@2) node[pos=.5,above,inner sep=1pt] {\ensuremath{#1}};}
\end{tikzpicture}\xspace\xspace
}

\NewDocumentCommand{\myto}{+O{} +O{0.5cm}}{%
\begin{tikzpicture}[baseline=-0.5ex] {
\node[inner sep=0](@1) at (0,0) {};
\node[inner sep=0](@2) at (#2,0) {};
\draw [arrows={-to}](@1)--(@2) node[pos=.5,above,inner sep=1pt] {\ensuremath{#1}};}
\end{tikzpicture}\xspace
}

\NewDocumentCommand{\paraarrow}{+O{} +O{0.4cm}}{%
\begin{tikzpicture}[baseline=-0.5ex] {
\node[inner sep=0](@1) at (0,0) {};
\node[inner sep=0](@2) at (#2,0) {};
\node[inner sep=0](@3) at (0.07,0) {};
\draw [arrows={-to}](@1)--(@2) node[pos=.5,above,inner sep=1pt] {\ensuremath{#1}};
\draw [arrows={-to}](@1)--(@3);}
\end{tikzpicture}\xspace
}

\NewDocumentCommand{\paradparrow}{+O{} +O{0.4cm}}{%
\begin{tikzpicture}[baseline=-0.5ex] {
\node[inner sep=0](@1) at (0,0) {};
\node[inner sep=0](@2) at (#2,0) {};
\node[inner sep=0](@3) at (0.07,0) {};
\draw [arrows={-Triangle[open]}](@1)--(@2) node[pos=.5,above,inner sep=1pt] {\ensuremath{#1}};
\draw [arrows={-to}](@1)--(@3);}
\end{tikzpicture}\xspace
}

\newcommand{\oset}[2]{%
  {\mathop{#2}\limits^{\vbox to 1\ex@{\kern-\tw@\ex@
   \hbox{\scriptsize #1}\vss}}}}

\newcommand{\osetthree}[2]{%
  {\mathop{#2}\limits^{\vbox to 3\ex@{\kern-\tw@\ex@
   \hbox{\scriptsize #1}\vss}}}}

\newcommand{\osetfive}[2]{%
  {\mathop{#2}\limits^{\vbox to 5\ex@{\kern-\tw@\ex@
   \hbox{\scriptsize #1}\vss}}}}

\newcommand{\osetminus}[2]{%
  {\mathop{#2}\limits^{\vbox to -2\ex@{\kern-\tw@\ex@
   \hbox{\scriptsize #1}\vss}}}}
\makeatother

\newcommand{\rootto}{\mathrel{\smash{\stackrel{\raisebox{3.4pt}{\scriptsize $\epsilon\:$}}{\smash{\rightarrow}}}}}

\newcommand{\fun}[1]{\mathrm{#1}}

\renewcommand{\phi}{\varphi}
\renewcommand{\emptyset}{\varnothing}

\newcommand{\pos}{\fun{pos}}

\newcommand{\posT}{\fun{pos}_{\SignatureA}}

\newcommand{\posDT}{\fun{pos}_{\SignatureD \cup \SignatureA}}

\newcommand{\subA}{\fun{ann}}

\newcommand{\anno}{\#} 

\newcommand{\VV}{\mathcal{V}}

\newcommand{\SignatureADC}{\Sigma^\#}

\newcommand{\SignatureC}{\mathcal{C}}
\newcommand{\SignatureD}{\mathcal{D}}
\newcommand{\SignatureA}{\mathcal{D}^\#}

\newcommand{\NN}{\mathbb{N}}

\newcommand{\tored}[3]{
  \mathrel{
    \xhookrightarrow{{}_{\scriptstyle #1}}
    \!\!{}^{#2}_{#3}
  }
}

\newcommand{\defemph}[1]{{\rm #1}}

\renewcommand{\epsilon}{\varepsilon}

%% file: abstract.tex
\begin{abstract}
  \emph{Dependency pairs}
    are one of the most powerful techniques for proving
    termination of term rewrite systems (TRSs), and they are used in almost all
    tools for termination analysis of TRSs.
    Problem \#106 of the RTA List of Open Problems asks for an adaption of
    dependency pairs for \emph{relative termination}. 
    Here, infinite rewrite sequences are allowed, but one wants to prove that a certain
    subset of the rewrite
    rules cannot be used infinitely often.
    Dependency pairs were recently adapted to 
    \emph{annotated dependency pairs (ADPs)} to prove almost-sure termination of probabilistic TRSs. 
    In this paper, we develop a novel adaption of ADPs for relative
    termination.
    We implemented our new ADP framework in our tool \aprove{} and
    evaluate it in comparison to 
    state-of-the-art tools for relative termination of TRSs.
\end{abstract}

%% file: introduction.tex
\section{Introduction}\label{Introduction}

Termination is an important topic in program verification.
There is a wealth of work on
automatic termination analysis of term rewrite systems (TRSs) which can also be used to
analyze termination of programs in many other languages.
Essentially all current termination tools for TRSs (e.g., \aprove~\cite{JAR-AProVE2017}, \natt~\cite{natt_sys_2014}, \muterm~\cite{gutierrez_mu-term_2020}, \ttttwo~\cite{ttt2_sys}, etc.)
use \emph{dependency pairs (DPs)}
\cite{arts2000termination,gieslLPAR04dpframework,giesl2006mechanizing,hirokawa2005automating,DBLP:journals/iandc/HirokawaM07}.

A combination of two TRSs $\R$ and $\R^{=}$
is considered to be ``\emph{relatively terminating}'' if there is no 
rewrite sequence that uses infinitely many steps
with rules from $\R$ (whereas rules from $\R^{=}$ may be used infinitely often).
Relative termination of TRSs has been
studied since decades \cite{DBLP:phd/dnb/Geser90}, and approaches based on
relative rewriting are used for many different applications, e.g.,
\cite{koprowskiZantema2005LivenessRelRew,fuhs2019DerivToRuntime,DBLP:journals/lmcs/NageleFZ17,DBLP:journals/corr/ZanklK14,DBLP:journals/jar/HirokawaM11,DBLP:conf/lpar/KleinH12,DBLP:conf/lopstr/IborraNV09,DBLP:journals/aaecc/NishidaV10,DBLP:conf/flops/Vidal08}.

However, while techniques and tools for analyzing ordinary termination of  TRSs
are very powerful due to the use of DPs, 
most approaches for automated analysis of relative termination are quite restricted in
power. Therefore, 
one of the largest open problems regarding DPs is Problem \#106 of the RTA List of Open Problems \cite{RTALoop}: 
\emph{Can we use the dependency pair method to prove relative termination?}
A first major step towards an answer to this question was presented in
\cite{iborra2017relative} by giving  criteria for $\R$ and $\R^{=}$ that allow the use of
ordinary DPs for relative termination. 

Recently, we adapted DPs  in order to analyze probabilistic innermost term rewriting, by using so-called
\emph{annotated dependency pairs (ADPs)} \cite{FLOPS2024}
or \emph{dependency tuples (DTs)} \cite{kassinggiesl2023iAST} (which were originally proposed
for innermost complexity analysis of  TRSs \cite{noschinski2013analyzing}).\footnote{As shown in
\cite{FLOPS2024}, using ADPs instead of DTs leads to a more elegant,
more powerful, and less complicated framework, and to completeness of
the underlying \emph{chain criterion}.}
In these adaptions, one \pagebreak[2] considers all \emph{defined} function symbols in the\linebreak[2]
right-hand side
of a rule at once, whereas ordinary DPs 
consider them separately.

In this paper, we show that considering the defined symbols on right-hand sides
separately (as for DPs) does not suffice for relative termination. On the other hand,
we do not need to consider all of them at once either.
Instead, we introduce a new definition of ADPs that is suitable for relative
termination and develop a corresponding ADP framework for automated relative
termination proofs of TRSs.
Moreover, while ADPs and DTs were only applicable for \emph{innermost}  rewriting in
\cite{FLOPS2024,kassinggiesl2023iAST,noschinski2013analyzing}, we now adapt ADPs to \emph{full}
(relative) rewriting, i.e., we do not impose any specific evaluation strategy.
So while \cite{iborra2017relative} presented conditions under which the \emph{ordinary
classical} DP framework can be used to 
prove relative termination, in this paper we develop the first \emph{specific} DP
framework for relative termination.

\medskip

\noindent
\textbf{Structure:} We start with preliminaries on relative rewriting in
\Cref{Relative Term Rewriting}.
In \Cref{DP Framework} we recapitulate the core processors of the DP framework.
Moreover, we state the main results of \cite{iborra2017relative} on using ordinary
DPs for relative termination.
Afterwards, we introduce our novel notion of \emph{annotated dependency pairs} for
relative termination in \Cref{Relative DP Framework}
and present a corresponding new ADP
framework in \Cref{Relative ADP Processors}.
We implemented our framework in the tool \aprove{} and
in  \Cref{Evaluation and Conclusion}, we evaluate our implementation in comparison to
other state-of-the-art  tools.
 All proofs can be found in \Cref{Appendix}.

%% file: reltrs.tex
\section{Relative Term Rewriting}\label{Relative Term Rewriting}

We assume familiarity with term rewriting \cite{baader_nipkow_1999} and regard (finite) TRSs
over a (finite) signature $\Sigma$ and a set of variables $\VSet$.

\begin{example}\label{ex:divlTRS}
Consider the following TRS $\R_{\tdivl}$, where $\tdivl(x,\xs)$ computes the
number that results from dividing $x$ by each element of the list $\xs$. 
As usual, natural numbers are 
represented by the function symbols $\O$ and $\ts$, and lists are represented via $\tnil$
and $\tcons$. 
Then $\tdivl(\ts^{24}(\O), \tcons(\ts^4(\O), \tcons(\ts^3(\O),\tnil)))$ evaluates to
$\ts^2(\O)$, because $(24/4)/3 = 2$. Here, $\ts^2(\O)$ stands for $\ts(\ts(\O))$, etc.

\vspace*{-0.5cm}

{\footnotesize
\hspace*{-0.85cm}
\begin{minipage}[t]{5.1cm}
    \begin{align}
        \label{R-minus-rule-2} \tminus(x,\O) &\to x\\
        \label{R-minus-rule-1} \tminus(\ts(x),\ts(y)) &\to \tminus(x,y) \\
        \label{R-div-rule-1} \tdiv(x,\ts(\O)) &\to x\!
    \end{align}
\end{minipage}\hspace{.3cm}
\begin{minipage}[t]{7.1cm}
    \begin{align}
        \label{R-div-rule-2} \tdiv(\ts(x),\ts(y)) &\to \ts(\tdiv(\tminus(x,y),\ts(y)))\\
        \label{R-list-rule-2} \tdivl(x,\tnil) &\to x \\
        \label{R-list-rule-1} \tdivl(x,\tcons(y,\xs)) &\to \tdivl(\tdiv(x,y),\xs) \!
         \end{align}
\end{minipage}}
\end{example}

\smallskip

\noindent 
A TRS $\R$ induces a \emph{rewrite relation} ${\to_{\R}} \subseteq \TSet{\Sigma}{\VSet}
\times \TSet{\Sigma}{\VSet}$ on terms where $s \to_{\R} t$ holds if there is a position
 $\pi \in \pos(s)$, 
a rule $\ell \to r \in \R$, and a substitution $\sigma$ such that $s|_{\pi}=\ell\sigma$ and $t = s[r\sigma]_{\pi}$.
For example, we have $\tminus(\ts(\O),\ts(\O)) \to_{\R_{\tdivl}} \tminus(\O,\O) \to_{\R_{\tdivl}} \O$.
We call a TRS $\R$ \emph{strongly normalizing (SN)} or \emph{terminating} if $\to_{\R}$ is
well founded. Using the DP framework, one can easily prove that
$\R_{\tdivl}$ is SN (see \Cref{Dependency Pairs for Ordinary Term Rewriting}). In
particular, in each application of the recursive $\tdivl$-rule 
\eqref{R-list-rule-1}, the length of the list in $\tdivl$'s second argument is
decreased by one.

In the relative setting, one considers two TRSs
$\R$ and $\R^{=}$.
We say that $\R$ is \emph{relatively strongly normalizing} w.r.t.\ $\R^{=}$ (i.e.,
$\R / \R^{=}$ is SN) if there is no infinite $(\to_{\R} \cup \to_{\R^{=}})$-rewrite sequence that uses an infinite number of $\to_{\R}$-steps.
We refer to $\R$ as the \emph{main} and $\R^{=}$ as the \emph{base} TRS.

\begin{example}\label{ex:mset1}
    For example, let $\R_{\tdivl}$ be the \emph{main} TRS.
    Since the order of the list elements does not affect \pagebreak[2] the termination
   of $\R_{\tdivl}$, 
     this algorithm also works for multisets.
   To abstract lists to multisets, we add the \emph{base} TRS $\R^{=}_{\tset} = \{\eqref{B-com-rule-1}\}$.
    \begin{equation}
        \label{B-com-rule-1} \tcons(x, \tcons(y,\zs)) \to \tcons(y, \tcons(x,\zs)) 
    \end{equation}
    $\R^{=}_{\tset}$ is non-terminating, since it can
    switch elements in a list arbitrarily often.
    How-\linebreak ever, $\R_{\tdivl} / \R^{=}_{\tset}$ is SN as each application of
    Rule \eqref{R-list-rule-1} still reduces the list length.
\end{example}

We will use the following four examples to show why a naive
adaption of dependency pairs does not work in the relative setting and why we need our new
notion of \emph{annotated dependency pairs}.
These examples represent different types of infinite rewrite sequences
that lead to non-termination in the relative setting: \emph{redex-duplicating},
\emph{redex-creating} (or ``-emitting''), and \emph{ordinary infinite sequences}.

\begin{example}[Redex-Duplicating]\label{example:redex-duplicating}
    Consider the TRSs $\R_1 = \{\ta \to \tb\}$ and $\R_1^{=} = \{\tf(x) \to \tc(\tf(x),x)\}$.
    $\R_1 / \R_1^{=}$ is not SN due to the infinite rewrite sequence $\underline{\tf(\ta)}
    \to_{\R_1^{=}}\linebreak \tc(\tf(\ta),\underline{\ta}) \to_{\R_1} \tc(\underline{\tf(\ta)},\tb) \to_{\R_1^{=}}
    \tc(\tc(\tf(\ta),\underline{\ta}),\tb)\to_{\R_1}
      \tc(\tc(\tf(\ta),\tb),\tb)\to_{\R_1^{=}}
    \ldots$\
    The reason is that $\R_1^{=}$ can be used to duplicate an arbitrary $\R_1$-redex infinitely often.
\end{example}

\begin{example}[Redex-Creating on Parallel Position]\label{example:redex-creating}
    Next, consider $\R_2 = \{\ta \to \tb\}$ and $\R_2^{=} = \{\tf \to \tc(\tf,\ta)\}$.
    $\R_2 / \R_2^{=}$ is not SN as we have the infinite rewrite sequence $\underline{\tf}
    \to_{\R_2^{=}} \tc(\tf,\underline{\ta}) \to_{\R_2} \tc(\underline{\tf},\tb) \to_{\R_2^{=}}
    \tc(\tc(\tf,\underline{\ta}),\tb) \to_{\R_2}
 \tc(\tc(\underline{\tf},\tb),\tb) \to_{\R_2^{=}} \ldots$\
 Here, $\R_2^{=}$ can create an
 $\R_2$-redex infinitely often (where
 in the right-hand side $\tc(\tf,\ta)$ of $\R_2^{=}$'s rule, the
 $\R_2^=$-redex $\tf$ and
 the created $\R_2$-redex $\ta$ are on parallel positions).
\end{example}

\begin{example}[Redex-Creating on Position Above]\label{example:redex-creatingAbove}
  Let $\R_3 = \{\ta(x) \to \tb(x)\}$ and $\R_3^{=} = \{\tf \to \ta(\tf)\}$.
    $\R_3 / \R_3^{=}$ is not SN as we have $\underline{\tf} \to_{\R_3^{=}}
    \underline{\ta}(\tf) \to_{\R_3} \tb(\underline{\tf}) \to_{\R_3^{=}}
    \tb(\underline{\ta}(\tf)) \to_{\R_3} 
 \tb(\tb(\underline{\tf})) \to_{\R_3^{=}}\ldots$, i.e., again
    $\R_3^{=}$ can be used to create an
 $\R_3$-redex infinitely often.
 In the right-hand side $\ta(\tf)$ of
$\R_3^=$'s rule,
the position  of the created $\R_3$-redex $\ta(\ldots)$
    is above the position of the  $\R_3^=$-redex $\tf$.
\end{example}

\begin{example}[Ordinary Infinite]\label{example:ordinary-infinite}
  Finally, consider $\R_4 = \{\ta \to \tb\}$ and $\R_4^{=} = \{ \tb \to \ta\}$.
Here, the base TRS $\R_4^{=}$ can neither duplicate nor create an $\R_4$-redex infinitely often,
    but in combination with the main TRS $\R_4$ we obtain the
  infinite rewrite sequence $\ta \to_{\R_4}
    \tb \to_{\R_4^{=}} \ta \to_{\R_4}
    \tb \to_{\R_4^{=}} \ldots$\ Thus,
    $\R_4 / \R_4^{=}$ is not SN. 
\end{example}

%% file: dp.tex
\section{DP Framework}\label{DP Framework}

We first recapitulate dependency pairs for ordinary (non-relative) rewriting in \Cref{Dependency Pairs for Ordinary Term Rewriting}
and summarize existing results on DPs for relative rewriting in \Cref{Dependency Pairs for Relative Termination}.

\subsection{Dependency Pairs for Ordinary Term Rewriting}\label{Dependency Pairs for Ordinary Term Rewriting}

We recapitulate DPs and the two most important processors of the DP framework, and refer to, e.g.,
\cite{arts2000termination,gieslLPAR04dpframework,giesl2006mechanizing,hirokawa2005automating,DBLP:journals/iandc/HirokawaM07}
for more details.
As an example, we show how to prove termination of $\R_{\tdivl}$ without the base $\R^{=}_{\tset}$.
We decompose the signature
$\Sigma =  \SignatureC \uplus  \SignatureD$ of a TRS $\R$ 
such that $f \in \SignatureD$ if $f = \rootsym(\ell)$ for some rule $\ell \to r \in \R$.
The symbols in $\SignatureC$ and $\SignatureD$ are called 
\emph{constructors} and \emph{defined symbols} of $\R$, respectively. 
For every $f \in \SignatureD$, we introduce a fresh \emph{annotated symbol} $f^{\#}$ of the same arity.
Let $\SignatureA$ denote the set of all annotated symbols, and $\SignatureADC = \Sigma \uplus \SignatureA$.
To ease readability, we often use capital letters like $\tF$ instead of $\tf^\#$.
For any term $t = f(t_1,\ldots,t_n) \in \TSet{\Sigma}{\VSet}$ with $f \in \SignatureD$, 
let $t^{\#} = f^{\#}(t_1,\ldots,t_n)$.
For a rule $\ell \to r$ and each subterm $t$ of $r$ with \pagebreak[2] defined root symbol, one obtains a
\emph{dependency pair} $\ell^\# \to t^\#$.
Let $\DPair{\R}$ denote the set of all dependency pairs of the TRS $\R$.

\begin{example}
  \label{example:dependency-pair}
    For $\R_{\tdivl}$ from \Cref{ex:divlTRS}, we obtain the following five dependency pairs.

    \vspace*{-.5cm}
    
    {\footnotesize
    \hspace*{-.9cm}
    \begin{minipage}[t]{5.7cm}
             \begin{align}
          \label{R-div-deppair-3} \tM(\ts(x),\ts(y)) &\to \tM(x,y) \\
            \label{R-div-deppair-2} \tD(\ts(x),\ts(y)) &\to \tM(x,y) \\
           \label{R-div-deppair-1} \tD(\ts(x),\ts(y)) &\to \tD(\tm(x,y),\ts(y))        
            \end{align}
    \end{minipage}\hspace*{.3cm}
    \begin{minipage}[t]{6.6cm}
        \begin{align}
            \label{R-div-deppair-4} \tDL(x,\tcons(y,\xs)) &\to \tD(x,y) \\
            \label{R-div-deppair-5} \tDL(x,\tcons(y,\xs)) &\to \tDL(\tdiv(x,y),\xs) 
        \end{align}
    \end{minipage}}
\end{example}

The DP framework operates on \emph{DP problems} $(\C{P}, \R)$ where
$\C{P}$ is a (finite) set of DPs, and $\R$ is a (finite) TRS. 
A (possibly infinite) sequence $t_0, t_1, t_2,
\ldots$ with $t_i \rootto_{\C{P}} \circ \to_{\R}^* t_{i+1}$ for all $i$ is a $(\C{P}, \R)$-\emph{chain}.
Here, $\rootto$ denotes rewrite steps at the root.
Intuitively, a chain represents subsequent ``function calls''  in evaluations. 
Between two function calls (corres\-ponding to steps with $\C{P}$, called $\mathbf{p}$-steps) one can evaluate the
arguments using arbitrary many steps with $\R$ (called $\mathbf{r}$-steps).
So $\mathbf{r}$-steps are rewrite steps that are needed in order to enable another
$\mathbf{p}$-step at a position above later on.\linebreak
For example, $\tDL(\ts(\O), \tcons(\ts(\O), \tnil)), \tDL(\ts(\O),\tnil)$ is a
$(\DPair{\R_{\tdivl}}, \R_{\tdivl})$-chain, as $\tDL(\ts(\O), \tcons(\ts(\O), \tnil))
\rootto_{\DPair{\R_{\tdivl}}} \tDL(\tdiv(\ts(\O),\ts(\O)),\tnil) \to_{\R_{\tdivl}}^*\!\tDL(\ts(\O),\tnil)$.

A DP problem $(\C{P}, \R)$ is called
\emph{strongly normalizing (SN)} if there is no infinite $(\C{P}, \R)$-chain.
The main result on DPs is the \emph{chain criterion} which states that a TRS
$\R$ is SN iff $(\DPair{\R}, \R)$ is SN.
The key idea of the DP framework is a \emph{divide-and-conquer} approach which
applies \emph{DP processors} to transform DP problems into simpler sub-problems.
A \emph{DP processor} $\Proc$ has the form $\Proc(\C{P}, \R) = \{(\C{P}_1,\R_1), \ldots, (\C{P}_n,\R_n)\}$, 
where $\C{P}, \C{P}_1, \ldots, \C{P}_n$ are sets of DPs and $\R, \R_1, \ldots, \R_n$ are TRSs. 
$\Proc$ is \emph{sound} if $(\C{P}, \R)$ is SN whenever 
$(\C{P}_i,\R_i)$ is SN for all $1 \leq i \leq n$. 
It is \emph{complete} if $(\C{P}_i,\R_i)$ is SN for all 
$1 \leq i \leq n$ whenever $(\C{P}, \R)$ is SN.

So for a TRS $\R$, one starts with the initial
DP problem $(\DPair{\R}, \R)$ and applies sound 
(and preferably complete) DP processors until all sub-problems are ``solved'' (i.e.,
DP processors transform them to the empty set).
This allows for modular ter-\linebreak mination
proofs, as different techniques can be applied on each sub-problem $(\C{P}_i, \R_i)$.

One of the most important processors is the \emph{dependency graph processor}.
The \emph{$(\C{P}, \R)$-dependency graph} indicates
which DPs can be used after each other in  
chains.\linebreak
Its nodes are $\C{P}$ and there is an edge from $s_1 \to t_1$ to $s_2 \to t_2$ if there
are substitutions
\begin{wrapfigure}[5]{r}{0.14\textwidth}
  \begin{center}
  \scriptsize
    \vspace*{-1.3cm}
    \begin{tikzpicture}
        \node[shape=rectangle,draw=black!100] (A) at (1,0.7) {\eqref{R-div-deppair-5}};
        \node[shape=rectangle,draw=black!100] (B) at (1,1.4) {\eqref{R-div-deppair-4}};
        \node[shape=rectangle,draw=black!100] (C) at (0,0) {\eqref{R-div-deppair-3}};
        \node[shape=rectangle,draw=black!100] (D) at (0,.7) {\eqref{R-div-deppair-2}};
        \node[shape=rectangle,draw=black!100] (E) at (0,1.4) {\eqref{R-div-deppair-1}};
   
        \path [->,in=290,out=250,looseness=5] (A) edge (A);
        \path [->] (A) edge (B);
        \path [->] (B) edge (D);
        \path [->] (B) edge (E);
        \path [->,in=20,out=340,looseness=5] (C) edge (C);
        \path [->] (D) edge (C);
        \path [->] (E) edge (D);
        \path [->,in=110,out=70,looseness=5] (E) edge (E);
    \end{tikzpicture}
    \caption*{}
  \end{center}
\end{wrapfigure}
$\sigma_1, \sigma_2$ with $t_1 \sigma_1 \to_{\R}^* s_2 \sigma_2$.
The $(\DPair{\R_{\tdivl}}, \R_{\tdivl})$-dependency graph is on the right.
Any infinite $(\C{P}, \R)$-chain corresponds to
an infinite path in the dependency graph, and since the graph is finite, this infinite
path must end in a strongly connected component (SCC).\footnote{Here, a
set $\C{P}'$ of dependency pairs is  an \emph{SCC} if it is a maximal cycle,
i.e., it is a maximal set such that for any $s_1 \to t_1$ and $s_2 \to
t_2$ in $\C{P}'$ there is
a non-empty path from $s_1 \to t_1$ to $s_2 \to
t_2$ which only traverses nodes from $\C{P}'$.}
Hence, it suffices to consider the SCCs of this graph independently.

\begin{restatable}[Dep.\ Graph Processor]{theorem}{depgraph}\label{DGP}
    For the SCCs $\C{P}_1, \ldots, \C{P}_n$ of the $(\C{P}, \R)$-dependency graph,  
    $\Proc_{\mathtt{DG}}(\C{P},\R) = \{(\C{P}_1,\R), \ldots, (\C{P}_n,\R)\}$ is sound and complete. 
\end{restatable}

While the exact dependency graph is not computable in general, there are 
techniques to over-approximate it automatically, see, e.g.,
\cite{arts2000termination,giesl2006mechanizing,hirokawa2005automating}.
In our example, $\Proc_{\mathtt{DG}}(\DPair{\R_{\tdivl}}, \R_{\tdivl})$ yields
$\bigl(\{\eqref{R-div-deppair-3}\}, \R_{\tdivl}\bigr)$,
$\bigl(\{\eqref{R-div-deppair-1}\}, \R_{\tdivl}\bigr)$,
 and $\bigl(\{\eqref{R-div-deppair-5}\}, \R_{\tdivl}\bigr)$.

The second crucial processor adapts classical reduction orders to DP problems.
A \emph{reduction pair} $(\succsim, \succ)$ consists of two \pagebreak[2]
relations on terms such that $\succsim$
is reflexive, transitive, and
closed under contexts and substitutions, and $\succ$ is a well-founded
order that is closed under substitutions but does
not have to be closed under contexts.
Moreover, $\succsim$ and $\succ$ must be compatible, 
i.e., ${\succsim} \circ {\succ} \circ {\succsim} \, \subseteq \, {\succ}$.
The \emph{reduction pair processor}
requires that all rules and dependency pairs are weakly decreasing,
and it removes those DPs that are strictly decreasing.

\begin{restatable}[Reduction Pair Processor]{theorem}{rpp}\label{RPP}
    Let $(\succsim, \succ)$ be a reduction pair such that $\C{P} \cup \R \subseteq \, \succsim$.
    Then $\Proc_{\mathtt{RPP}}(\C{P},\R) = \{(\C{P} \, \setminus \succ, \R)\}$ is sound and complete.
\end{restatable}

For example, one can use reduction pairs based on
polynomial interpretations \cite{lankford1979proving}.
A \emph{polynomial interpretation} $\Pol$ is a $\SignatureADC$-algebra which maps every
function symbol $f \in \SignatureADC$ to a polynomial $f_{\Pol} \in \IN[\VSet]$.
$\Pol(t)$ denotes the \emph{interpretation} of a term $t$ by the $\SignatureADC$-algebra $\Pol$.
Then $\Pol$ induces a reduction pair
$(\succsim, \succ)$ where $t_1 \succsim t_2$ ($t_1 \succ t_2$) holds if the inequation $\Pol(t_1) \geq \Pol(t_2)$
($\Pol(t_1) > \Pol(t_2)$) is true for all
instantiations of its variables by natural numbers.

For the three remaining
DP problems in our example, we can apply
the reduction pair processor using
the polynomial interpretation
which maps $\O$ to $0$, $\ts(x)$ to $x + 1$,
$\tcons(y,\xs)$ to $\xs + 1$, $\tDL(x,\xs)$ to $\xs$,
and all other symbols to their first arguments. 
Since $\eqref{R-div-deppair-3}$, $\eqref{R-div-deppair-1}$,
and $\eqref{R-div-deppair-5}$ are strictly decreasing, 
$\Proc_{\mathtt{RPP}}$ transforms all three remaining DP problems 
into DP problems of the form $(\emptyset, \ldots)$. 
As $\Proc_{\mathtt{DG}}(\emptyset, \ldots) = \emptyset$ 
and all processors used are sound, this means that there is no
infinite chain for the initial DP problem 
$(\DPair{\R_{\tdivl}}, \R_{\tdivl})$ and thus, $\R_{\tdivl}$ is SN.

\subsection{Dependency Pairs for Relative Termination}\label{Dependency Pairs for Relative Termination}

Up to now, we only considered DPs for ordinary termination of TRSs.
The easiest idea to use DPs in the relative setting is to start with the DP problem 
$(\DPair{\R \cup \R^{=}}, \R \cup \R^{=})$.
This would prove termination of $\R \cup \R^{=}$, which implies termination of $\R / \R^{=}$, but
ignores that the rules in $\R^{=}$ do not have to terminate.
Since termination of DP problems is already defined via a relative condition (finite chains
can only have finitely
many $\mathbf{p}$-steps but may have
infinitely many $\mathbf{r}$-steps),
another idea 
for proving termination of $\R / \R^{=}$ is to start
with the DP problem $(\DPair{\R}, \R \cup \R^{=})$, which only considers
the DPs of $\R$.
However,  this is unsound in general.

\begin{example}\label{ex:dps-dont-work-in-relative}
     The only defined symbol of $\R_2$  from \Cref{example:redex-creating} is $\ta$.
     Since the right-hand side of $\R_2$'s rule does not contain 
  defined symbols, we would get the DP problem
  $(\emptyset,\linebreak \R_2 \cup \R_2^{=})$, which is SN as it has no
  DP.
    Thus, we would falsely conclude that $\R_2 / \R_2^{=}$\linebreak is SN.
Similarly, 
this approach would also falsely ``prove'' SN for
\Cref{example:redex-duplicating,example:redex-creatingAbove}.
\end{example}

In \cite{iborra2017relative}, it was shown that under
certain conditions on $\R$ and $\R^{=}$, starting with the DP problem $(\DPair{\R \cup \R_a^{=}},
\R \cup \R^{=})$ for a subset $\R_a^{=} \subseteq \R^{=}$
is sound
for relative termination.\footnote{As before,
for the construction of $\DPair{\R
  \cup \R_a^{=}}$,  only the root symbols of left-hand sides of $\R
\cup \R_a^{=}$
are considered to be ``defined''.}
The two restrictions on the TRSs are \emph{dominance} and being \emph{non-duplicating}.
We say that $\R$ \emph{dominates} $\R^{=}$ if defined symbols of $\R$
do not occur in
the right-hand sides of rules of $\R^{=}$.
A TRS is \emph{non-duplicating} if no variable occurs more often
on the right-hand side of a rule than on its left-hand side.

\begin{restatable}[First Main Result of~\cite{iborra2017relative}, Sound and Complete]{theorem}{main-relative-rewrite-corollary-yamada-1}\label{theorem:main-relative-rewrite-corollary-yamada-1}
    Let $\R$ and $\R^{=}$ be TRSs such that $\R^{=}$ is non-duplicating and \pagebreak[2] $\R$ dominates $\R^{=}$.
    Then the DP problem
      $(\DPair{\R}, \R \cup \R^{=})$ is SN iff $\R / \R^{=}$ is SN.
\end{restatable}

\begin{restatable}[Second Main Result of~\cite{iborra2017relative}, only Sound]{theorem}{main-relative-rewrite-corollary-yamada-2}\label{theorem:main-relative-rewrite-corollary-yamada-2}
    Let $\R$ and $\R^{=} = \R^{=}_a \uplus \R^{=}_b$ be TRSs.
	If $\R^{=}_b$ is non-duplicating, $\R \cup \R^{=}_a$ dominates $\R^{=}_b$,
    and the DP problem $(\DPair{\R \cup \R^{=}_a}, \R \cup \R^{=})$ is SN, 
    then $\R / \R^{=}$ is SN.
\end{restatable}

\begin{example}
  For the main TRS $\R_{\tdivl}$ from \Cref{ex:divlTRS} and base TRS $\R^{=}_{\tset}$ from \Cref{ex:mset1}\linebreak
  we can apply
    \Cref{theorem:main-relative-rewrite-corollary-yamada-1} and consider the DP problem
    $(\DPair{\R_{\tdivl}}, \R_{\tdivl} \cup \R^{=}_{\tset})$, since $\R^{=}_{\tset}$ is
    non-duplicating and $\R_{\tdivl}$ dominates $\R^{=}_{\tset}$.
    As for $(\DPair{\R_{\tdivl}}, \R_{\tdivl})$, the DP
    framework can prove that $(\DPair{\R_{\tdivl}}, \R_{\tdivl} \cup
    \R^{=}_{\tset})$ is SN. In this way, the tool \natt{} 
    which implements the results of
    \cite{iborra2017relative}  proves that $\R_{\tdivl} / \R^{=}_{\tset}$ is SN.
In contrast,  a direct application of  simplification orders fails to prove
    SN for $\R_{\tdivl} / \R^{=}_{\tset}$ because simplification orders already fail to
    prove termination of $\R_{\tdivl}$.
\end{example}

\begin{example}\label{ex:mainExample}
    If we consider $\R^{=}_{\tset 2}$ with the rule 
    \begin{equation}
        \label{B-com-rule-2} \tdivl(z,\tcons(x, \tcons(y,\zs))) \to \tdivl(z,\tcons(y, \tcons(x,\zs)))
    \end{equation}
    instead of $\R^{=}_{\tset}$ as the base TRS, then $\R_{\tdivl} / \R^{=}_{\tset 2}$ remains strongly normalizing,
    but we cannot use \Cref{theorem:main-relative-rewrite-corollary-yamada-1} since $\R_{\tdivl}$ does not
    dominate $\R^{=}_{\tset 2}$. If we try to split $\R^{=}_{\tset 2}$ as in
    \Cref{theorem:main-relative-rewrite-corollary-yamada-2}, then
    $\emptyset \neq \R^{=}_a \subseteq \R^{=}_{\tset 2}$ implies $\R^{=}_a = \R^{=}_{\tset
    2}$, but
    $\R^{=}_{\tset  2}$ is 
    non-terminating.
    Therefore, all previous tools for relative termination fail in proving that $\R_{\tdivl} /
    \R^{=}_{\tset 2}$ is SN.
    In \Cref{Relative DP Framework} we will present our novel DP framework which can prove
    relative termination of relative TRSs 
    like $\R_{\tdivl} / \R^{=}_{\tset 2}$.
\end{example}

As remarked in \cite{iborra2017relative},
\Cref{theorem:main-relative-rewrite-corollary-yamada-1,theorem:main-relative-rewrite-corollary-yamada-2} 
are unsound if one only considers \emph{minimal} chains, i.e., if
for a DP problem $(\C{P},\R)$ one only considers 
chains $t_0, t_1, \ldots$, where all\linebreak $t_i$ are $\R$-strongly normalizing.
In the DP framework for ordinary rewriting,
the re\-striction to minimal chains allows the use of further processors, e.g., based on
\emph{usable}\linebreak \emph{rules} \cite{giesl2006mechanizing,DBLP:journals/iandc/HirokawaM07} or the \emph{subterm criterion} \cite{DBLP:journals/iandc/HirokawaM07}.
As shown in \cite{iborra2017relative}, usable rules and the subterm
criterion can nevertheless be 
applied if $\R^{=}$ is \emph{quasi-terminating} \cite{dershowitz_termination_1987}, 
i.e., the\linebreak set $\{t \mid s \to_{\R^{=}}^* t \}$ is finite for every term $s$.
This restriction would also be needed to integrate processors that rely on
minimality
into our new framework in \Cref{Relative DP Framework}.

%% file: reldp.tex
\section{Annotated Dependency Pairs for Relative Termination}\label{Relative DP Framework}

As shown in \Cref{Dependency Pairs for Relative Termination},
up to now there only exist criteria \cite{iborra2017relative} that state
when it is sound to apply
\emph{ordinary} DPs for proving relative termination, but there is no \emph{specific} DP-based technique
to analyze relative termination directly. To solve this problem, we now
adapt the concept of \emph{annotated dependency pairs} (ADPs) for relative termination.
ADPs were introduced in \cite{FLOPS2024} to prove innermost almost-sure termination of probabilistic term rewriting.
In the relative setting, we can use similar dependency pairs as in the probabilistic setting,
but with a different rewrite relation $\tored{}{}{}$ to deal with non-innermost rewrite steps.
Compared to \cite{iborra2017relative},
we can (a) remove the requirement of dominance, which will be handled
by the dependency graph processor, and (b) allow for ADP processors that are specifically designed for the relative
setting before possibly moving to ordinary DPs.
The requirement that 
$\R^{=}$ must be non-duplicating remains, since DPs do not help in analyzing
redex-duplicating sequences as in \Cref{example:redex-duplicating}, where the crucial redex
$\ta$ is not generated from a ``function call'' \pagebreak[2]
in the right-hand side of a rule,
but it just corresponds to a duplicated variable.
To handle TRSs $\R / \R^{=}$ where $\R^{=}_{dup} \subseteq\R^{=}$
is duplicating,
one can move the duplicating rules to the main TRS $\R$ and 
try to prove relative termination of
$(\R \cup \R^{=}_{dup})/(\R^{=} \setminus \R^{=}_{dup})$ instead,
or one can try to find a reduction pair $(\succsim, \succ)$ where
$\succ$ is closed under contexts such that $\R \cup \R^= \subseteq {\succsim}$ and
$\R^{=}_{dup} \subseteq {\succ}$. Then it suffices to prove relative termination of $(\R
\setminus \succ, \R^{=} \setminus \succ)$ instead.

For ordinary termination, we create a separate DP for each occurrence of a
defined symbol in the right-hand side of a rule (and no DP is created for rules without
defined symbols in their right-hand sides).
This would work to detect
\emph{ordinary infinite} sequences
like the one in \Cref{example:ordinary-infinite}
in the relative setting, i.e., such an infinite sequence would give rise to an infinite chain. 
However,  as shown in \Cref{ex:dps-dont-work-in-relative}, this
would not suffice to detect infinite redex-creating sequences as in
\Cref{example:redex-creating}
with   $\R_2 = \{\ta \to \tb\}$ and $\R_2^{=} = \{\tf \to \tc(\tf,\ta)\}$:
$\underline{\tf}
    \to_{\R_2^{=}} \tc(\tf,\underline{\ta}) \to_{\R_2} \tc(\underline{\tf},\tb) \to_{\R_2^{=}}
    \tc(\tc(\tf,\underline{\ta}),\tb) \to_{\R_2}
     \ldots$\

Here, (1) we need a DP for
the rule $\ta \to \tb$ to detect the reduction of the created
$\R_2$-redex $\ta$, although $\tb$ is a constructor.
Moreover,
(2) both defined symbols $\tf$ and $\ta$ in the right-hand side of
$\tf \to \tc(\tf, \ta)$ have to be considered simultaneously:
We need $\tf$ to create an infinite number of $\R_2$-redexes, and we need $\ta$ since it
is the created $\R_2$-redex.
Hence, for rules from the base TRS $\R_2^=$, we have to consider all possible pairs of
defined symbols in their right-hand sides simultaneously.\footnote{For relative
termination, it suffices to
consider \emph{pairs} of defined symbols.
The reason is that to ``track'' a non-terminating reduction, one only has
to
consider a single redex plus possibly another redex of the base
TRS which may later create a redex again.
}
This is not needed for the main TRS $\R_2$, i.e., if the $\tf$-rule were in the
main TRS, then the $\tf$ in the right-hand side could be considered separately from the $\ta$ that it generates.
Therefore, we distinguish between \emph{main} and \emph{base ADPs} (that are
generated from the main and the base TRS, respectively).

As in \cite{FLOPS2024},  we now annotate defined
symbols directly in the original rewrite rule instead of extracting annotated subterms
from its right-hand side. In this way, we may have terms containing several annotated
symbols, which 
allows us to consider pairs of defined symbols in right-hand sides
simultaneously.

\begin{definition}[Annotations]
    For $t \in \TSet{\SignatureADC}{\VSet}$ 
    and $\Sigma' \subseteq \SignatureADC \cup \VSet$, let $\pos_{\Sigma'}(t)$ be the
    set of all positions of $t$
    with symbols or variables from $\Sigma'$.
    For $\Phi \subseteq \posDT(t)$,
    $\anno_\Phi(t)$ is the variant of $t$ where the symbols at positions from $\Phi$
    are annotated and all\linebreak other annotations are removed.
    Thus, $\posT(\anno_\Phi(t)) = \Phi$, and
    $\anno_\emptyset(t)$ removes all an-\linebreak
    notations from $t$, where we often write
    $\flat(t)$ instead of $\anno_\emptyset(t)$.
    Moreover, for a singleton $\{\pi \}$, we often write $\anno_\pi$ instead of $\anno_{\{\pi\}}$.
   We write $t \trianglelefteq_{\#}^\pi s$ if
there is a $\pi \in \posT(s)$ and $t = \flat(s|_\pi)$
(i.e., $t$ results from  a subterm of $s$ with annotated root
symbol by removing its annotations). We also write $\trianglelefteq_{\#}$ instead of $\trianglelefteq_{\#}^\pi$.
\end{definition}

\begin{example}
    If $\tf \in \SignatureD$, then we have $\anno_{1}(\tf(\tf(x))) =
    \anno_{1}(\tF(\tF(x))) = \tf(\tF(x))$ and $\flat(\tF(\tF(x))) = \tf(\tf(x))$. 
    Moreover, we have $\tf(x) \trianglelefteq_{\#}^{1} \tf(\tF(x))$.
\end{example}

While in \cite{FLOPS2024} all defined symbols on the right-hand sides of rules were
annotated, 
we now define our novel variant of \emph{annotated dependency pairs}
for relative rewriting.

\begin{definition}[Annotated Dependency Pair]\label{def:Canonical-ADPs}
\hspace*{-.2cm}    A rule $\ell\!\to\!r$ with $\ell\!\in\!\TSet{\Sigma}{\VSet} \setminus\linebreak
    \VSet$, $r \in \TSet{\SignatureADC}{\VSet}$, and $\VSet(r) \subseteq \VSet(\ell)$ is
    called an
    \defemph{annotated dependency pair (ADP)}.
 
    Let $\SignatureD$ be the defined symbols of $\R \cup \R^=$, and for $n \in
    \NN$, let $\ADPair{n}{\ell \to r}  =  
    \{\ell \to\linebreak \anno_{\Phi}(r) \mid \Phi \subseteq \pos_{\SignatureD}(r), |\Phi| \leq n\}$.
\pagebreak[2]  The  \defemph{canonical main ADPs} for $\R$ are $\ADPairMain{\R} = \bigcup\limits_{\ell \to r \in \R}
  \!\!\!\!  \ADPairMain{\ell\!\to\!r}$ and  the \defemph{canonical base ADPs} for $\R^=\!$
are  $\ADPairBase{\R^{=}}\!= \!\!\!
    \bigcup\limits_{\ell \to r \in \R^=} \!\!\!\! \ADPairBase{\ell\!\to\!r}$.
\end{definition}

So the left-hand side of an ADP is just the left-hand side of the original rule.
The right-hand side  results from the right-hand side of the original rule
by replacing certain defined symbols $f$ with $f^{\#}$.
Whenever
we have two ADPs  $\ell \to \anno_{\Phi'}(r)$,
$\ell \to \anno_{\Phi}(r)$
with  $\Phi' \subset \Phi$, then
we only consider $\ell \to \anno_{\Phi}(r)$ and remove
 $\ell \to \anno_{\Phi'}(r)$.

\begin{example}\label{ADP-Divl}
    The canonical ADPs
    of \Cref{example:redex-creating} are $\ADPairMain{\R_2} = \{ \ta
    \to \tb\}$
    and $\ADPairBase{\R_2^{=}} = \{\tf \to
    \tc(\tF,\tA)\}$ and for
    \Cref{example:redex-creatingAbove} we get $\ADPairMain{\R_3} = \{ \ta(x)
    \to \tb(x)\}$
    and $\ADPairBase{\R_3^{=}} = \{\tf \to
    \tA(\tF)\}$.
    For $\R_{\tdivl}/\R^{=}_{\tset 2}$ from \Cref{ex:divlTRS,ex:mainExample},
    the ADPs $\ADPairMain{\R_{\tdivl}}$ are

\vspace*{-.4cm}
    
    {\footnotesize
    \hspace*{-.7cm}\begin{minipage}[t]{5.1cm}
        \begin{align}
            \label{R-div-adp-2} \tminus(x,\O) &\to x\\
            \label{R-div-adp-1} \tminus(\ts(x),\ts(y)) &\to \tM(x,y) \\
            \label{R-div-adp-3} \tdiv(x,\ts(\O)) &\to x\\
            \label{R-div-adp-4} \tdivl(x,\tnil) &\to x \!
        \end{align}
    \end{minipage}
    \begin{minipage}[t]{7.5cm}
        \begin{align}
            \label{R-div-adp-5} \tdiv(\ts(x),\ts(y)) &\to \ts(\tD(\tminus(x,y),\ts(y)))\\
            \label{R-div-adp-6} \tdiv(\ts(x),\ts(y)) &\to \ts(\tdiv(\tM(x,y),\ts(y)))\\
            \label{R-div-adp-7} \tdivl(x,\tcons(y,\xs)) &\to \tDL(\tdiv(x,y),\xs) \\
            \label{R-div-adp-8} \tdivl(x,\tcons(y,\xs)) &\to \tdivl(\tD(x,y),\xs) \!
        \end{align}
    \end{minipage}}
    
    \vspace*{-.1cm}

    \begin{align}
      \label{B-com-adp-2}
      \hspace*{-.2cm}
         \text{and $\ADPairBase{\R^{=}_{\tset 2}}$ contains 
      {\small $\tdivl(z, \tcons(x, \tcons(y,\zs))) \to \tDL(z, \tcons(y, \tcons(x,\zs)))$}}
    \end{align}
\end{example}

In \cite{FLOPS2024},
ADPs were only used for innermost rewriting.
We now modify their rewrite relation and define what happens 
with annotations inside the substitutions during a rewrite step.
To simulate redex-creating sequences as in \Cref{example:redex-creatingAbove}
with ADPs (where the position of the created redex $\ta(\ldots)$
is above the position of the creating redex $\tf$),
ADPs should be able to rewrite above annotated arguments
without removing their annotation (we will demonstrate that in
\Cref{ex:ADPs-for-redex-creation-2}).
Thus, for an ADP $\ell \to\linebreak r$ with $\ell|_\pi = x$, we use a 
\emph{variable reposition function (VRF)} to indicate which occur-\linebreak rence of $x$ in $r$ should
keep the annotations if one rewrites an instance of $\ell$ where the subterm at position
$\pi$ is annotated.
So a VRF maps  positions of variables in the left-hand side of a rule to
positions of the same variable in the right-hand side.

\begin{definition}[Variable Reposition Function]\label{def:Var-Repos-Func}
    Let $\ell \to r$ be an ADP.
	A function $\varphi: \pos_{\VSet}(\ell) \to \pos_{\VSet}(r) \cup \{\bot\}$ is called a
    \defemph{variable reposition function (VRF)} for
    $\ell \to r$
    iff
    $\ell|_\pi = r|_{\varphi(\pi)}$ whenever
 $\varphi(\pi) \neq \bot$.
\end{definition}

\begin{example}\label{example:rel-var-repos-function}
    For the ADP $\ta(x) \to \tb(x)$ for $\R_3$ from \Cref{example:redex-creatingAbove},
    if $x$ on position 1 of the left-hand side is instantiated by $\tF$,
    then the VRF $\varphi(1) = 1$ 
    indicates that
    this ADP rewrites $\tA(\tF)$ to $\tb(\tF)$, whereas
    $\varphi(1) = \bot$  means that
    it rewrites $\tA(\tF)$ to $\tb(\tf)$.
\end{example}

With VRFs we can define the rewrite relation for ADPs w.r.t.\ full rewriting.

\begin{definition}[$\tored{}{}{\C{P}}$]\label{def:ADP-Rewriting}
    Let $\C{P}$ be a set of ADPs.
    A term $s \in \TSet{\SignatureADC}{\VSet}$ rewrites to $t$ using $\C{P}$
    (denoted $s \tored{}{}{\C{P}} t$)
if there is a rule $\ell \to r \in \C{P}$, 
    a substitution $\sigma$, a position $\pi \in \posDT(s)$
    such that $\flat(s|_\pi) = \ell\sigma$, a VRF $\varphi$ for $\ell \to r$,
    and\footnote{\label{ADPComparison2}In \cite{FLOPS2024} there
    were two additional cases in the definition of the corresponding rewrite relation. One of
    them was needed for processors that restrict the set of rules applicable for
    $\mathbf{r}$-steps (e.g., based on usable rules), and the other case 
    was needed to
    ensure that the innermost
    evaluation strategy is not affected by the application of ADP processors. This is
    unnecessary here since we consider full rewriting. On the other hand,
     VRFs are new compared to
\cite{FLOPS2024}, since they are not needed for innermost rewriting.}
    \begin{equation*}
        \begin{array}{rll@{\quad}ll@{\qquad}l}
        t &=                  &s[\anno_{\Phi}(r\sigma)]_{\pi}  & 
        \text{if} & \pi \in \posT(s)    & (\mathbf{pr})\\ 
        t &=                  &s[\anno_{\Psi}(r\sigma)]_{\pi}  & 
        \text{if} & \pi \in\pos_\SignatureD(s) & (\mathbf{r})\!
        \end{array}
    \end{equation*}
    Here, $\Psi \!=\! \{\varphi(\rho).\tau \mid \rho \!\in\! \pos_{\VSet}(\ell), \,
    \varphi(\rho) \neq \bot,  \, \rho.\tau \!\in\! \posT(s|_{\pi}) \}$
    and $\Phi = \posT(r) \cup \Psi$.
\end{definition}
So $\Psi$ considers all positions of annotated symbols in $s|_{\pi}$ that
are below
positions $\rho$ of\linebreak variables in $\ell$. If the VRF maps $\rho$ to a variable position
$\rho'$ in
$r$, then the annotations below $\pi.\rho$ in $s$ are kept in the resulting subterm at
position $\pi.\rho'$
after the rewriting.

Rewriting with $\C{P}$ is like ordinary term rewriting, while considering and
modifying
annotations.
Note that we represent all DPs resulting from a rule as well as the original
rule by just one ADP.  
So the ADP $\tdiv(\ts(x),\ts(y)) \to \ts(\tD(\tminus(x,y),\ts(y)))$
represents both the DP resulting from $\tdiv$ in the right-hand side
of the rule \eqref{R-div-rule-2}, and the rule \eqref{R-div-rule-2} itself 
(by simply disregarding all annotations of the ADP).

Similar to the classical DP framework, our goal is to track specific reduction
sequences. As before, 
there are $\mathbf{p}$-steps where
a DP is applied at the
position of an annotated symbol. These steps may
introduce new annotations. Moreover, 
between two $\mathbf{p}$-steps there can be
several $\mathbf{r}$-steps.

A step of the form $(\mathbf{pr})$ at position $\pi$ in \Cref{def:ADP-Rewriting} 
represents  
a $\mathbf{p}$- or an  
$\mathbf{r}$-step\linebreak (or both), where an  
$\mathbf{r}$-step is only possible 
if one later rewrites an annotated symbol at a position above $\pi$.
All annotations are kept during this step except for annotations of subterms
that correspond to variables of the applied rule. Here, the used VRF $\varphi$ determines which of
these annotations are kept and which are removed.
As an example,
with the canonical ADP $\ta(x) \to \tb(x)$  from
$\ADPairMain{\R_3}$ we can rewrite
$\tA(\tF) \tored{}{}{\ADPairMain{\R_3}}  \tb(\tF)$ as in \Cref{example:rel-var-repos-function}.
Here, we have $\pi =
\varepsilon$, $\flat(s|_\varepsilon) = \ta(\tf) = \ell \sigma$, $r =
\tb(x)$, and the VRF $\varphi$ with 
$\varphi(1) = 1$ such that the annotation of $\tF$ in $\tA$'s argument is
kept in the  argument of $\tb$.

A step of the form $(\mathbf{r})$
rewrites at the position of a non-annotated defined symbol,
and represents just an $\mathbf{r}$-step.  
Hence, we remove all annotations
from the right-hand side $r$ of the ADP.
However, we may have to keep the annotations inside the substitution,
hence we move them according to the VRF.
For example, we obtain the rewrite step
$\ts(\tD(\underline{\tminus(\ts(\O),\ts(\O))},\ts(\O))) \tored{}{}{\ADPairMain{\R_\tdivl}}\ts(\tD(\tminus(\O,\O),\ts(\O)))$
using the ADP $\tminus(\ts(x),\ts(y)) 
\to \tM(x,y) \;$ \eqref{R-div-adp-1}
and any VRF.

A \emph{(relative) ADP problem} has the form
$(\C{P},\C{P}^{=})$, where $\C{P}$ and $\C{P}^{=}$ are finite sets of ADPs and $\C{P}^{=}$
is non-duplicating. 
$\C{P}$ is the set of all main ADPs and $\C{P}^{=}$ is the set of all base ADPs.
Now we can define chains in the relative setting.

\begin{definition}[Chains and Terminating ADP Problems]\label{def:relative-rewrite-chain}
    Let $(\C{P},\C{P}^{=})$ be an ADP problem.
    A sequence of terms $t_0, t_1, \ldots$ is a
    $(\C{P},\C{P}^{=})$-\defemph{chain} if we have $t_i \tored{}{}{\C{P} \cup \C{P}^{=}}
    t_{i+1}$ for all $i \in \IN$.
    The chain is called \defemph{infinite} if infinitely
    many of these rewrite steps use $\tored{}{}{\C{P}}$
    with Case $(\mathbf{pr})$. 
    We say that an ADP problem $(\C{P},\C{P}^{=})$ is \defemph{strongly normalizing (SN)} if
    there is no infinite $(\C{P},\C{P}^{=})$-chain.
\end{definition}

Note the two different forms of relativity in \Cref{def:relative-rewrite-chain}:
In a finite chain,
we may not only use infinitely many steps with $\C{P}^{=}$ but also infinitely many steps
with $\C{P}$ where Case $(\mathbf{r})$ applies.  
Thus, an ADP problem $(\C{P},\C{P}^{=})$
without annotated symbols or without any main ADPs (i.e., where $\C{P} = \emptyset$) is obviously SN.
Finally, we obtain our desired chain criterion.

\begin{restatable}[Chain Criterion for Relative Rewriting]{theorem}{RelChainCriterion}\label{theorem:relative-chain-crit}
    Let $\R$ and $\R^{=}$ be TRSs such that $\R^{=}$ is non-duplicating.
    Then $\R / \R^{=}$ is SN iff the ADP problem $(\ADPairMain{\R}, \ADPairBase{\R^{=}})$ is SN.
\end{restatable}

\begin{example}\label{ex:ADPs-for-redex-creation-1}
    The infinite rewrite sequence of \Cref{example:redex-creating} can be simulated by
    the following infinite chain  using $\ADPairMain{\R_2} = \{ \ta
    \to \tb\}$
    and $\ADPairBase{\R_2^{=}} = \{\tf \to
    \tc(\tF,\tA)\}$.
    \[\underline{\tF}
    \tored{}{}{\ADPairBase{\R_2^{=}}} \tc(\tF,\underline{\tA}) \tored{}{}{\ADPairMain{\R_2}} \tc(\underline{\tF},\tb)
    \tored{}{}{\ADPairBase{\R_2^{=}}} \tc(\tc(\tF,\underline{\tA}),\tb) \tored{}{}{\ADPairMain{\R_2}}
    \ldots\]

    The steps with $\tored{}{}{\ADPairBase{\R_2^{=}}}$ use Case
    ($\mathbf{pr}$) at the position of the annotated symbol $\tF$
    and the steps 
    with $\tored{}{}{\ADPairMain{\R_2}}$ use ($\mathbf{pr}$) as well.
    For this infinite chain, we indeed need 
    two annotated symbols in the right-hand side of the base ADP: If $\tA$ were not annotated (i.e., if we had the ADP
    $\tf \to
    \tc(\tF,\ta)$), then the step with 
    $\tored{}{}{\ADPairMain{\R_2}}$ would just use Case ($\mathbf{r}$) and the chain would
    not be considered ``infinite''. If $\tF$ were not annotated
    (i.e., if we had the ADP
    $\tf \to
    \tc(\tf,\tA)$), then we would have the step
     $\tf
    \tored{}{}{\ADPairBase{\R_2^{=}}} \tc(\tf,\ta)$ which uses
    Case ($\mathbf{r}$)
    and removes all
    annotations from the right-hand side. Hence, again the chain would not be considered ``infinite''.
\end{example}

\begin{example}\label{ex:ADPs-for-redex-creation-2}
  The infinite rewrite sequence of
  \Cref{example:redex-creatingAbove} is simulated by the following chain with
   $\ADPairMain{\R_3} = \{ \ta(x) \to \tb(x)\}$
    and $\ADPairBase{\R_3^{=}} = \{\tf \to \tA(\tF)\}$.
    \[\underline{\tF}
    \tored{}{}{\ADPairBase{\R_3^{=}}} \underline{\tA}(\tF) \tored{}{}{\ADPairMain{\R_3}} \tb(\underline{\tF})
    \tored{}{}{\ADPairBase{\R_3^{=}}} \tb(\underline{\tA}(\tF)) \tored{}{}{\ADPairMain{\R_3}}\tb(\tb(\underline{\tF}))  \tored{}{}{\ADPairBase{\R_3^{=}}}
    \ldots\]
    Here, it is important to use the VRF $\varphi(1) = 1$ for $\ta(x) \to \tb(x)$
    which keeps the annotation of $\tA$'s argument
    $\tF$ during the rewrite steps with $\ADPairMain{\R_3}$, i.e., these steps must yield
    $\tb(\tF)$ instead of $\tb(\tf)$    
    to generate further  subterms  $\tA(\ldots)$ afterwards.
\end{example}

%% file: ADPframework.tex
\section{The Relative ADP Framework}\label{Relative ADP Processors}

Now we present processors for our novel
relative ADP framework.
An \emph{ADP processor}\linebreak $\Proc$ has the form
$\Proc(\C{P},\C{P}^{=}) = \{(\C{P}_1,\C{P}^{=}_1), \ldots,
(\C{P}_n,\C{P}^{=}_n)\}$,  
where $\C{P}, \C{P}_1, \ldots, \C{P}_n,\linebreak \C{P}_1^{=}, \ldots, \C{P}_n^{=}$ are sets of ADPs. 
 $\Proc$ is \emph{sound} if $(\C{P},\C{P}^{=})$ is SN whenever 
$(\C{P}_i,\C{P}^{=}_i)$ is SN for all $1 \leq i \leq n$. 
It is \emph{complete} if
$(\C{P}_i,\C{P}^{=}_i)$ is SN for all 
$1 \leq i \leq n$ whenever $(\C{P},\C{P}^{=})$ is SN.
To prove relative termination of $\R/\R^=$, we start with the canonical ADP problem $(\ADPairMain{\R},\ADPairBase{\R^{=}})$
and apply sound 
(and preferably complete) ADP processors until all sub-problems are  transformed to the empty set.

In \Cref{Derelatifying Processors}, we
present two processors to remove (base) ADPs, and
in \Cref{Relative Dependency Graph Processor,Relative Reduction Pair Processor}, we adapt the main processors of the classical DP framework from \Cref{Dependency Pairs for Ordinary Term Rewriting}
to the relative setting.
As mentioned,  the soundness and completeness proofs for our processors and the chain criterion (\Cref{theorem:relative-chain-crit})
can be found in \Cref{Appendix}.

\subsection{Derelatifying Processors}\label{Derelatifying Processors}

The following two \emph{derelatifying} processors can be used to switch from ADPs to ordinary DPs,
similar to \Cref{theorem:main-relative-rewrite-corollary-yamada-1,theorem:main-relative-rewrite-corollary-yamada-2}.
We extend $\flat$ to ADPs and sets of ADPs $\SSS$
by defining $\flat(\ell  \to r) = \ell \to \flat(r)$
and $\flat(\SSS) = \{\ell  \to \flat(r) \mid \ell \to r \in \SSS\}$.

If the ADPs in $\C{P}^{=}$ contain no annotations anymore,
then it suffices to use ordinary DPs.
The corresponding set of DPs for a set of ADPs $\C{P}$ is defined as 
$\DP{\C{P}} = \{\ell^\# \to t^\# \mid \ell \to r \in \C{P}, t \trianglelefteq_{\#} r\}$.

\begin{restatable}[Derelatifying Processor (1)]{theorem}{DerelProcOne}\label{theorem:derel-proc-1}
     Let $(\C{P}, \C{P}^{=})$ be an ADP problem such that $\flat(\C{P}^{=}) = \C{P}^{=}$.    
     Then $\Proc_{\mathtt{DRP1}}(\C{P}, \C{P}^{=}) = \emptyset$ is sound and complete
     iff the ordinary DP problem
     $(\DP{\C{P}}, \flat(\C{P} \cup \C{P}^{=}))$ \pagebreak[2] is SN.
\end{restatable}

Furthermore,  similar to \Cref{theorem:main-relative-rewrite-corollary-yamada-2},
we can always move ADPs from $\C{P}^{=}$ to $\C{P}$,
but such a processor is only sound and not complete.
However, it may help to satisfy 
the requirements of \Cref{theorem:derel-proc-1} by moving ADPs with 
annotations from $\C{P}^{=}$ to $\C{P}$ such that
the ordinary DP framework can be used  afterwards.

\begin{restatable}[Derelatifying Processor (2)]{theorem}{DerelProcTwo}\label{theorem:derel-proc-2}
    Let $(\C{P}, \C{P}^{=})$ be an ADP problem, and let $\C{P}^{=} = \C{P}^{=}_{a} \uplus \C{P}^{=}_{b}$.
    Then $\Proc_{\mathtt{DRP2}}(\C{P}, \C{P}^{=}) = \{(\C{P} \cup \mathtt{split}(\C{P}^{=}_{a}),
    \C{P}^{=}_{b})\}$ is sound.
    Here, $\mathtt{split}(\C{P}^{=}_{a}) = \{\ell \to \anno_{\pi}(r) \mid \ell \to r \in \C{P}^{=}_{a}, \pi \in \fun{pos}_{\SignatureD^\#}(r)\}$.    
\end{restatable}
\noindent
So if $\C{P}^{=}_{a}$ contains an ADP with two annotations, then we split it into two
ADPs, where each only contains
 a single annotation.

\begin{example}\label{terminatingRedexDuplCreate}
    There are also examples that are redex-creating and terminating, e.g., $\R_2 = \{ \ta \to \tb \}$ 
    and the base TRS $\R_2^{='} =
    \{ \tf(\ts(y)) \to \tc(\tf(y),\ta) \}$.
    Relative (and full) termination of this example can easily be
    shown by using
 the second derelatifying processor from \Cref{theorem:derel-proc-2} to
    replace the base ADP
    $\tf(\ts(y)) \to \tc(\tF(y),\tA)$ by the main ADPs $\tf(\ts(y)) \to \tc(\tF(y),\ta)$ and
    $\tf(\ts(y)) \to \tc(\tf(y),\tA)$. Then one can use the processor of
    \Cref{theorem:derel-proc-1} to switch to the ordinary DPs
$\tF(\ts(y)) \to \tF(y)$ and $\tF(\ts(y)) \to \tA$.
 \end{example}
  
\subsection{Relative Dependency Graph Processor}\label{Relative Dependency Graph Processor}

Next, we develop a dependency graph processor in the relative setting.
The definition of the dependency graph is analogous to the one in the standard setting and
thus, the same techniques can be used to over-approximate it automatically.

\begin{definition}[Relative Dependency Graph]\label{def:rel-dependency-graph}
    Let $(\C{P}, \C{P}^{=})$ be an ADP problem. 
    The $(\C{P}, \C{P}^{=})$-\defemph{dependency graph}
    has the nodes $\C{P} \cup \C{P}^{=}$ and there is an edge from
    $\ell_1 \to r_1$ to $\ell_2 \to r_2$ 
    if there exist substitutions $\sigma_1, \sigma_2$ and a term $t \trianglelefteq_{\#} r_1$ 
    such that $t^\# \sigma_1 \rightarrow_{\flat(\C{P} \cup \C{P}^{=})}^*
  \ell_2^\# \sigma_2$.
\end{definition}

So similar to the standard dependency graph,
there is an edge from an ADP $\ell_1 \to r_1$ to
$\ell_2 \to r_2$  if the rules of 
$\flat(\C{P} \cup \C{P}^{=})$ (without annotations) can reduce an instance of a
subterm $t$ of $r_1$ to an instance of $\ell_2$, if one only annotates
the roots of $t$ and $\ell_2$ (i.e., then the rules can only be applied below the root). 

\textcolor{white}{.}
\begin{wrapfigure}[5]{r}{0.42\textwidth}
            \scriptsize
        \vspace*{-1cm}
     \begin{tikzpicture}
            \node[shape=rectangle,draw=black!100] (A) at (0,1) {\eqref{R-div-adp-1}};
            \node[shape=rectangle,draw=black!100] (B) at (0,0) {\eqref{R-div-adp-2}};
            \node[shape=rectangle,draw=black!100] (C) at (1,0) {\eqref{R-div-adp-3}};
            \node[shape=rectangle,draw=black!100] (D) at (2,0) {\eqref{R-div-adp-4}};
            \node[shape=rectangle,draw=black!100] (E) at (2,1) {\eqref{R-div-adp-5}};
            \node[shape=rectangle,draw=black!100] (F) at (1,1) {\eqref{R-div-adp-6}};
            \node[shape=rectangle,draw=black!100] (G) at (3,1) {\eqref{R-div-adp-7}};
            \node[shape=rectangle,draw=black!100] (H) at (3,0) {\eqref{R-div-adp-8}};
            \node[shape=circle,draw=black!100] (I) at (4,0.5) {\eqref{B-com-adp-2}};
        
            \path [->,in=110,out=70,looseness=5] (A) edge (A);
            \path [->] (A) edge (B);
            \path [->] (F) edge (A);
            \path [->] (F) edge (B);
            \path [->,in=110,out=70,looseness=5] (E) edge (E);
            \path [->] (E) edge (C);
            \path [->] (E) edge (F);
            \path [->,in=110,out=70,looseness=5] (G) edge (G);
            \path [->] (G) edge (D);
            \path [->] (G) edge (H);
            \path [->] (G) edge (I);
            \path [->,in=330,out=210,looseness=1] (H) edge (C);
            \path [->] (H) edge (E);
            \path [->] (H) edge (F);
            \path [->,in=330,out=30,looseness=5] (I) edge (I);
            \path [->] (I) edge (G);
            \path [->] (I) edge (H);
        \end{tikzpicture}
  \end{wrapfigure}

\vspace*{-.8cm}
\begin{example}\label{ex:rel-dependency-graph}
    The dependency graph for the ADP problem $(\ADPairMain{\R_{\tdivl}},
    \ADPairBase{\R^{=}_{\tset 2}})$
    from \Cref{ADP-Divl} is shown on the right. 
    Here, nodes from $\ADPairMain{\R_{\tdivl}}$ are denoted by rectangles and the
    node from $\ADPairBase{\R^{=}_{\tset 2}}$ is a circle.
\end{example}

To detect possible ordinary infinite rewrite sequences as in \Cref{example:ordinary-infinite},
we again have to regard SCCs of the dependency graph,
where we only need to consider SCCs that contain a node from $\C{P}$,
because otherwise, all steps in the SCC are relative.
However, in the relative ADP framework,
non-termination can also be due to chains representing redex-creating sequences.
Here,
it does not suffice to
look at SCCs.
Thus, the relative dependency graph processor differs substantially from the corresponding
processor for ordinary rewriting (and also from the corresponding processor for the
probabilistic ADP framework in \cite{FLOPS2024}). 

\begin{example}[Dependency Graph for Redex-Creating TRSs]\label{ex:DepGraphRedexCreating}
    For $\R_2$ and $\R_2^=$ from \Cref{example:redex-creating},
    the dependency graph for  $(\ADPairMain{\R_2}, \ADPairBase{\R_2^{=}})$ from
    \Cref{ex:ADPs-for-redex-creation-1} can be \pagebreak[2] seen on the

    \noindent
    \begin{minipage}[t]{9cm}
        right.
        Here, we cannot regard the SCC $\{\tf \to \tc(\tF,\tA)\}$ separately,
        as we need the rule  $\ta \to \tb$ from $\ADPairMain{\R_2}$ \linebreak
    \end{minipage}
    \hspace*{0.05cm}
    \begin{minipage}[t]{2.5cm}
        \begin{center}
            \scriptsize
            \vspace*{-0.4cm}
            \begin{tikzpicture}
                \node[shape=rectangle,draw=black!100] (A) at (0,0) {$\ta \to \tb$};
                \node[shape=rectangle,draw=black!100] (B) at (2,0) {$\tf \to \tc(\tF,\tA)$};
            
                \path [->,in=110,out=70,looseness=5] (B) edge (B);
                \path [->] (B) edge (A);
            \end{tikzpicture}
        \end{center}
    \end{minipage}

    \vspace*{-0.3cm}
    \noindent
    to reduce the created redex.
    To find the ADPs that can reduce the created redexes,
    we have to regard the outgoing paths
    from the SCCs of $\C{P}^=$  to ADPs of $\C{P}$. 
\end{example}

The structure that we are looking for in the redex-creating case is 
 a path from an SCC to a node from $\C{P}$
 (i.e., a form of a \emph{lasso}),
which is \emph{minimal} in the sense that if we reach a node from $\C{P}$, then we stop
and do not move further along the edges of the graph.
Moreover, the SCC needs to contain an ADP with more than one annotated symbol, as otherwise the
generation of the infinitely many
$\C{P}$-redexes would not be possible.
Here, it suffices to look at SCCs in the graph restricted to only $\C{P}^{=}$-nodes (i.e.,
to SCCs in the $(\flat(\C{P}),\C{P}^{=})$-dependency graph).
The reason is that
if the SCC contains a node from $\C{P}$, then as mentioned above,
we have to prove anyway
that the SCC does not give rise to infinite chains.

\begin{definition}[$\mathtt{SCC}^{(\C{P},\C{P}^{=})}_{\C{P}'}$, $\mathtt{Lasso}$]\label{def:lasso}
    Let $(\C{P},\C{P}^{=})$ be an ADP problem.
    For any $\C{P}' \subseteq \C{P} \cup \C{P}^{=}$,
    let $\mathtt{SCC}^{(\C{P},\C{P}^{=})}_{\C{P}'}$ denote 
    the set of all SCCs of the
    $(\C{P},\C{P}^{=})$-dependency graph that contain an ADP from
    $\C{P}'$. Moreover, let  $\C{P}^{=}_{>1} \subseteq \C{P}^{=}$ denote the set of all
    ADPs from $\C{P}^{=}$ with
    more than one annotation.
    Then the set of all \defemph{minimal lassos} is defined as
    $\mathtt{Lasso} = \{\QQ \cup \{n_1, \ldots, n_k\} \mid \QQ \in
    \mathtt{SCC}^{(\flat(\C{P}),\C{P}^{=})}_{\C{P}^{=}_{>1}}, \; n_1,\ldots,n_k$ is a path such that $n_1 \in \QQ, \; n_k \in \C{P}, \text{ and } n_i \not\in \C{P} \text{ for all } 1 \leq i \leq k-1\}$.
\end{definition}

We remove the annotations of ADPs which do not have to be considered anymore\linebreak for
$\mathbf{p}$-steps due to the dependency graph, 
but we keep the ADPs for possible $\mathbf{r}$-steps and thus, consider
them as relative (base) ADPs.

\begin{restatable}[Dep.\ Graph Processor]{theorem}{RelativeDepGraphProc}\label{theorem:rel-DGP}
    Let $(\C{P},\C{P}^{=})$ be an ADP problem.
    Then

    \vspace*{-.4cm}
    
    {\small\begin{align*}
        \Proc_{\mathtt{DG}}(\C{P},\C{P}^{=}) & =
        \{ (\,\C{P} \cap \QQ, \;
        (\C{P}^= \cap \QQ)\cup        \flat( \,(\C{P} \cup \C{P}^{=}) \setminus \QQ \,)\,
           )
        \mid \QQ
        \in \mathtt{SCC}_{\C{P}}^{(\C{P}, \C{P}^{=})} \cup \mathtt{Lasso}\} \! 
    \end{align*}
    }

    \vspace*{-.1cm}

    \noindent
    is sound and complete.   
\end{restatable}

\begin{example}\label{ex:DivlDepGraph}
    For $(\ADPairMain{\R_{\tdivl}}, \ADPairBase{\R^{=}_{\tset 2}})$ from
    \Cref{ex:rel-dependency-graph} we have three SCCs 
    $\{\eqref{R-div-adp-1}\}$, $\{\eqref{R-div-adp-5}\}$, 
    and $\{\eqref{R-div-adp-7},\eqref{B-com-adp-2}\}$ containing nodes from $\ADPairMain{\R_{\tdivl}}$.
    The set $\{\eqref{B-com-adp-2}\}$ is the only
    SCC of $(\flat(\ADPairMain{\R_{\tdivl}}), \ADPairBase{\R^{=}_{\tset 2}})$ and there
    are paths from that SCC to the ADPs $\eqref{R-div-adp-7}$ and
    $\eqref{R-div-adp-8}$ of $\C{P}$. However, they are not in
    $\mathtt{Lasso}$, because the SCC $\{\eqref{B-com-adp-2}\}$ does not contain an ADP with more than one
    annotation.
    Hence, we result in the three new ADP problems 
    $(\{\eqref{R-div-adp-1}\} \cup \flat(\ADPairMain{\R_{\tdivl}} \setminus \{\eqref{R-div-adp-1}\}), \{\flat(\ref{B-com-adp-2})\})$,
    $(\{\eqref{R-div-adp-5}\} \cup \flat(\ADPairMain{\R_{\tdivl}} \setminus \{\eqref{R-div-adp-5}\}), \{\flat(\ref{B-com-adp-2})\})$,
    and 
    $(\{\eqref{R-div-adp-7}\} \cup \flat(\ADPairMain{\R_{\tdivl}} \setminus \{\eqref{R-div-adp-7}\}),\{(\ref{B-com-adp-2})\})$.
    For the first two of these new ADP problems, 
    we can use the derelatifying processor of \Cref{theorem:derel-proc-1} and prove SN via ordinary DPs, since their base
    ADP $\flat(\ref{B-com-adp-2})$ does not contain any annotated symbols anymore.
\end{example}

The dependency graph processor in combination with the derelatifying processors of \Cref{theorem:derel-proc-1,theorem:derel-proc-2}
already subsumes the techniques of
\Cref{theorem:main-relative-rewrite-corollary-yamada-1,theorem:main-relative-rewrite-corollary-yamada-2}.
The reason is that if $\R$ dominates $\R^{=}$, then there is no edge from an ADP of $\ADPairBase{\R^{=}}$ to
any ADP of $\ADPairMain{\R}$ in the $(\ADPairMain{\R}, \ADPairBase{\R^{=}})$-dependency
graph. Hence, there are no minimal lassos and the
dependency graph processor just creates ADP problems from the SCCs of $\ADPairMain{\R}$
where the base ADPs do not have any annotations anymore. Then \Cref{theorem:derel-proc-1}
allows us to switch to ordinary DPs.
For example, if we consider $\R^{=}_{\tset}$ instead of $\R^{=}_{\tset 2}$, then the
dependency graph processor \pagebreak[2]  only yields the two subproblems for the SCCs 
$\{\eqref{R-div-adp-1}\}$ and $\{\eqref{R-div-adp-5}\}$, 
where the base ADPs do not\linebreak contain any annotations anymore.
Then, we can move to ordinary
DPs via \Cref{theorem:derel-proc-1}.

Compared to
\Cref{theorem:main-relative-rewrite-corollary-yamada-1,theorem:main-relative-rewrite-corollary-yamada-2},
the dependency graph allows for more precise
over-approximations than just ``dominance'' in order
to detect when the base ADPs do not depend on 
the main ADPs.  Moreover,
the derelatifying processors of \Cref{theorem:derel-proc-1,theorem:derel-proc-2}
allow us to switch to  the ordinary DP framework also for subproblems which result 
from the application of other processors of our relative ADP framework.
In other words, \Cref{theorem:derel-proc-1,theorem:derel-proc-2} allow us to apply this
switch in a modular way, even if their prerequisites
do not hold for the initial canonical ADP problem (i.e., even if the prerequisites of \Cref{theorem:main-relative-rewrite-corollary-yamada-1,theorem:main-relative-rewrite-corollary-yamada-2}
do not hold for the whole TRSs).

\subsection{Relative Reduction Pair Processor}\label{Relative Reduction Pair Processor}

Next, we adapt the reduction pair processor to ADPs for relative rewriting.
While the reduction pair processor for ADPs in the probabilistic setting
\cite{FLOPS2024} was restricted to polynomial interpretations,
we now allow arbitrary
reduction pairs
using a similar idea as in 
the reduction pair processor from \cite{noschinski2013analyzing} for complexity analysis
via dependency tuples.

To find out which ADPs cannot be used for infinitely many $\mathbf{p}$-steps,
the idea is not to compare the annotated left-hand side with the
whole right-hand side, but just with the set of its annotated subterms.
To combine these subterms in the case of
ADPs with two or no annotated symbols, we extend the signature by two fresh \emph{compound} symbols
$\Com{0}$ and $\Com{2}$ of arity $0$ and $2$, respectively.
Similar to \cite{noschinski2013analyzing},
we have to use $\Com{}$\emph{-monotonic} and $\Com{}$\emph{-invariant} reduction pairs.

\begin{definition}[$\Com{}$-Monotonic, $\Com{}$-Invariant]\label{def:poly-interpretation-for-depset}
    For $r \in \TSet{\SignatureADC}{\VSet}$, we define
    $\subA(r) = \Com{0}$ if $r$ does not contain any annotation,
    $\subA(r) = t^\#$ if $t \trianglelefteq_{\#} r$ and $r$ only contains one annotated symbol,
    and $\subA(r) = \Com{2}(r_1^\#, r_2^\#)$ if $r_1 \trianglelefteq_{\#}^{\pi_1} r$,
    $r_2 \trianglelefteq_{\#}^{\pi_2} r$, 
    and $\pi_1 <_{lex} \pi_2$ where $<_{lex}$ is the (total)
    lexicographic order on positions.

    A reduction pair $(\succsim, \succ)$ is called \defemph{$\Com{}$-monotonic}
    if $\Com{2}(s_1, t) \succ
    \Com{2}(s_2, t)$ and $\Com{2}(t, s_1) \succ
    \Com{2}(t, s_2)$ for all $s_1,s_2,t \in \TSet{\SignatureADC}{\VSet}$ with $s_1 \succ s_2$.
    Moreover, it is  \defemph{$\Com{}$-invariant}
    if $\Com{2}(x,y) \sim \Com{2}(y,x)$ and
    $\Com{2}(x,\Com{2}(y,z)) \sim 
    \Com{2}(\Com{2}(x,y),z)$
    for ${\sim} = {\succsim} \cap {\precsim}$. 
\end{definition}
So for example, reduction pairs based on polynomial interpretations are
$\Com{}$-monotonic and $\Com{}$-invariant if $\Com{2}(x,y)$ is interpreted by $x + y$.

For an ADP problem $(\C{P},\C{P}^{=})$, 
now the reduction pair processor has to orient the non-annotated rules $\flat(\C{P} \cup
\C{P}^{=})$ weakly and for all ADPs $\ell \to r$,
it compares the annotated left-hand side  $\ell^\#$ with
$\subA(r)$. In strictly decreasing ADPs, one can then remove all annotations and consider
them as relative (base) ADPs again.

\begin{restatable}[Reduction Pair Processor]{theorem}{RelRPP}\label{thm:RelRPP}
    Let $(\C{P},\C{P}^{=})$ be an ADP problem and let
    $(\succsim, \succ)$ be a $\Com{}$-monotonic and $\Com{}$-invariant reduction pair
    such that $\flat(\C{P} \cup \C{P}^{=})\linebreak \subseteq {\succsim}$ and
    $\ell^\# \succsim \subA(r)$ for all $\ell \to r \in \C{P} \cup \C{P}^{=}$.
    Moreover, let $\PP_{\succ} \subseteq \C{P} \cup \C{P}^{=}$ 
    such that\linebreak $\ell^\# \succ \subA(r)$ for all $\ell \to r \in \PP_{\succ}$. 
    Then $\Proc_{\mathtt{RPP}}(\C{P},\C{P}^{=}) = \{(\C{P} \setminus \PP_{\succ}, (\C{P}^{=} \setminus \PP_{\succ}) \cup
    \flat(\PP_{\succ}))\}$ is sound and complete.
\end{restatable}

\begin{example}\label{example:rel-RPP}
    For the remaining ADP problem
    $(\{\eqref{R-div-adp-7}\} \cup \flat(\ADPairMain{\R_{\tdivl}} \setminus \{\eqref{R-div-adp-7}\}),\{(\ref{B-com-adp-2})\})$ 
    from \Cref{ex:DivlDepGraph}, we can apply the reduction pair processor
    using the polynomial interpretation from the end of \Cref{Dependency Pairs for Ordinary Term Rewriting} which maps 
    $\O$ to $0$, $\ts(x)$ to $x + 1$,
    $\tcons(y,\xs)$ to $\xs + 1$, $\tDL(x,\xs)$ to $\xs$,
    and all other symbols to their first arguments. 
    Then, $\eqref{R-div-adp-7}$ is oriented strictly (i.e., it is in
    $\PP_{\succ}$) and $\eqref{B-com-adp-2}$ is oriented weakly.
    Hence, we remove the annotation from $\eqref{R-div-adp-7}$ and move it to the base
    ADPs.
    Now there is no SCC with a main ADP anymore in the
    dependency graph, and thus the dependency graph processor 
    returns $\emptyset$.
    This proves SN for $(\ADPairMain{\R_{\tdivl}}, \ADPairBase{\R^{=}_{\tset 2}})$, hence
    $\R_{\tdivl} / \R^{=}_{\tset 2}$ is also SN.
\end{example}

\begin{example}\label{ex:RPPCreating}
    Regard the ADPs
    $\ta \to \tb$  and $\tf \to \tc(\tF,\tA)$ for
    the redex-creating  \Cref{example:redex-creating} again.
    When using  a polynomial interpretation  $\mathrm{Pol}$ that maps $\Com{0}$ to $0$ and
    $\Com{2}(x,y)$ to $x + y$, then for the reduction pair processor 
    one has to satisfy $\mathrm{Pol}(\tA) \geq 0$ and
    $\mathrm{Pol}(\tF) \geq \mathrm{Pol}(\tF) + \mathrm{Pol}(\tA)$, i.e.,
    one cannot
    make any of the ADPs strictly decreasing.

    In contrast, for the variant
    with the terminating base rule  $\tf(\ts(y)) \to \tc(\tf(y),\ta)$ from \Cref{terminatingRedexDuplCreate},
    we have the ADPs $\ta \to \tb$  and $\tf(\ts(y)) \to \tc(\tF(y),\tA)$. 
    Here, the second constraint is  $\mathrm{Pol}(\tF(\ts(y))) \geq
    \mathrm{Pol}(\tF(y)) + \mathrm{Pol}(\tA)$. To make 
    one of the ADPs strictly decreasing,
    one can set
    $\mathrm{Pol}(\tF(x)) = x$, $\mathrm{Pol}(\ts(x)) = x+1$, and
    $\mathrm{Pol}(\tA) = 1$ or $\mathrm{Pol}(\tA) = 0$.
    Then the reduction pair processor
    removes the annotations from the strictly decreasing ADP and 
     the dependency graph processor  proves SN.
\end{example}

%% file: evaluation.tex
\section{Evaluation and Conclusion}\label{Evaluation and Conclusion}

In this paper, we introduced the first
notion of (annotated) dependency pairs
and the first DP framework
for relative termination, which also features suitable
dependency graph and reduction pair processors for relative ADPs.
Of course, 
further classical DP processors can be adapted to our relative ADP framework as well. For
example, in our implementation of the novel ADP framework in our tool \aprove{} \cite{JAR-AProVE2017},
we also included a
straightforward adaption of the classical \emph{rule removal processor}
\cite{gieslLPAR04dpframework}, see 
\Cref{Appendix}.\footnote{This processor works
analogously to the preprocessing
at the beginning
of \Cref{Relative DP Framework}
which can be used to remove duplicating rules: For an  ADP problem $(\C{P},\C{P}^{=})$, it tries to find
a reduction pair 
$(\succsim, \succ)$ 
where $\succ$ is closed under contexts
such that $\flat(\C{P} \cup \C{P}^{=}) \subseteq {\succsim}$.
Then for $\PP_{\succ} \subseteq \C{P} \cup \C{P}^{=}$ 
with $\flat(\C{P}_{\succ}) \subseteq {\succ}$, the processor replaces the ADP by 
$(\C{P} \setminus \PP_{\succ}, \C{P}^{=} \setminus \PP_{\succ})$.}
In future work, we will investigate how to use our new form of 
ADPs for full (instead of innermost) rewriting also in the probabilistic setting 
and for complexity analysis.

To evaluate the new relative ADP framework, we compared its implementation in 
``\emph{new} \aprove{}'' 
to all other tools that participated
in the most recent \emph{termination competition (TermComp 2023)} \cite{termcomp}
on relative rewriting, i.e.,
\natt{} \cite{natt_sys_2014},  \ttttwo{} \cite{ttt2_sys}, \mnm{} \cite{FSCD19}, and ``\emph{old} \aprove{}'' which did
not yet contain the contributions of the current paper.
In \emph{TermComp 2023}, 
98 benchmarks were used for relative termination. However, these benchmarks only consist
of
examples where the main TRS
$\R$ dominates the base TRS $\R^{=}$ (i.e., which can be handled by \Cref{theorem:main-relative-rewrite-corollary-yamada-1} from
\cite{iborra2017relative})
or which can already
be solved via simplification orders directly.
Therefore, we extended the collection by
17 new examples,
including both
$\R_{\tdivl}/\R^{=}_{\tset}$ from \Cref{ex:divlTRS,ex:mset1},
and our leading example $\R_{\tdivl} / \R^{=}_{\tset 2}$ 
from \Cref{ex:mainExample} (where only \emph{new} \aprove{} can prove SN).
Except for $\R_{\tdivl}/\R^{=}_{\tset}$, in these examples
$\R$ does not dominate $\R^{=}$. 
Most of these examples adapt well-known classical TRSs from the
\emph{Termination Problem Data Base} \cite{tpdb} used at \emph{TermComp}
to the relative setting.
In the following table,
the number in the ``YES'' (``NO'') row indicates for how many of the 115
examples the respective
tool could prove (disprove) relative termination and ``MAYBE'' refers to the benchmarks
where the tool could not solve the problem within the timeout of 300~s per example. The numbers in
brackets are the respective results when only considering our new 17 examples.
``AVG(s)'' gives the average runtime of the tool
on solved examples in seconds.

\begin{center}
{\small    \begin{tabular}{||c | c | c | c | c | c||}
     \hline
      & \emph{new} \aprove & \natt & \emph{old} \aprove & \ttttwo & \mnm  \\ [0.5ex] 
     \hline
     YES & 78 (17) & 65 (7) & 47 (4) & 39 (3) & 0 (0)  \\ 
     \hline
     NO & 13 (0)& 5 (0)& 13 (0)& 7 (0)& 13 (0)\\ 
     \hline
     MAYBE & 24 (0)& 45 (10)& 55 (13)& 69 (14) & 102 (17) \\ 
     \hline
      AVG(s) & 6.68 & 0.38 & 3.67 & 1.61 & 1.28  \\ 
     \hline
    \end{tabular}}
\end{center}

The table clearly shows that while \emph{old} \aprove{} was already the second most powerful
tool for relative termination, the integration of the ADP framework in \emph{new} \aprove{}
yields a substantial advance in power (i.e., it only fails on 24 of the examples, compared
to 45 and 55 failures of \natt{} and \emph{old} \aprove, respectively).
In particular,  previous tools (including 
\emph{old} \aprove{}) often have problems with
relative TRSs where the main TRS does
 not dominate the base TRS, whereas the ADP framework can handle
 such examples.

A special form of relative TRSs are \emph{relative string rewrite systems (SRSs)}, where all function
symbols have arity 1.
Due to the base ADPs with two annotated
symbols on the right-hand side, 
here the ADP framework is less powerful
than dedicated techniques for string rewriting.
For the 403 relative SRSs at \emph{TermComp 2023},
the ADP framework only finds 71 proofs, mostly due to the dependency graph and the rule removal processor, 
while termination analysis via \aprove's
standard strategy for relative SRSs
succeeds on 209 examples, and the two most powerful tools for relative SRSs at
\emph{TermComp 2023} (\mnm{} and \matchbox{} \cite{matchbox})
succeed on 274 and 269 examples, respectively.

Another special form of relative rewriting is \emph{equational rewriting}, where one has
a set of equations $E$ which correspond to relative rules that can be applied in both directions.
In \cite{RTA01}, DPs were adapted to equational rewriting.
However, this approach requires
$E$-unification to be decidable and finitary
(i.e., for (certain) pairs of terms, 
it has to compute  finite complete sets of $E$-unifiers).
This works well if $E$
are AC- or C-axioms, and for this special case, dedicated techniques like
\cite{RTA01} are more powerful than our new ADP framework for relative termination.
For example, on the 76 AC- and C-benchmarks  for equational rewriting at \emph{TermComp 2023},
the  relative ADP framework  finds 36 proofs, while dedicated
tools for AC-rewriting like \aprove's equational strategy or \muterm{} \cite{gutierrez_mu-term_2020} succeed on
66 and 64 examples, respectively.
However, in general, the requirement of a finitary
$E$-unification algorithm is a  hard restriction.
In contrast to existing tools for equational rewriting,
our new ADP framework can be used for arbitrary
(non-duplicating) relative rules.

For details on our experiments, our collection of examples,
and for instructions on how to run our implementation
in \textsf{AProVE} via its \emph{web interface} or locally, see
\url{https://aprove-developers.github.io/RelativeDTFramework/}

%% file: main.bbl
\providecommand{\noopsort}[1]{}

%% file: appendix.tex
\section{Appendix}\label{Appendix}
In this appendix, we give all proofs for our new results and observations, and present an
additional rule removal processor for our ADP framework (\Cref{thm:RelRRP}) that results
from a straightforward adaption of the corresponding processor from the classical DP
framework \cite{gieslLPAR04dpframework}.

Before we start with the proof of the chain criterion,
for any infinite rewrite sequence $t_0 \to_{\R\cup \R^{=}} t_1 \to_{\R\cup \R^{=}} \ldots$
we define \emph{origin graphs} 
that indicate
which subterm of $t_{i+1}$ ``originates'' from which subterm of $t_i$. Therefore, these
graphs also indicate how annotations can move in the chains that correspond to this
rewrite sequence.
Note that due to the choice of the VRF and due to the fact that at most two defined symbols
are annotated in the right-hand sides of ADPs,
there are multiple possibilities 
for the annotations to move in a chain.
Each origin graph corresponds to one possible way that the annotations can move.

\begin{definition}[Origin Graph]\label{def:orig}
    Let $\R$ and $\R^{=}$ be two TRSs 
    and let $\Theta: t_0 \to_{\R \cup \R^{=}} t_1 \to_{\R \cup \R^{=}} \ldots$ be a rewrite sequence.
    A graph with the nodes $(i,\pi)$ for all $i \in \IN$ and all $\pi \in \pos(t_i)$
    is called an \defemph{origin graph} for $\Theta$
    if the edges are defined as follows:
    For $i \in \IN$,
    let
the rewrite step  $t_i \to_{\R \cup \R^{=}} t_{i+1}$ be performed using the rule
$\ell \to r \in \R\cup \R^{=}$, the position $\tau$, and the substitution $\sigma$,
i.e., $t_i|_\tau = \ell \sigma$ and $t_{i+1} = t_i[r \sigma]_\tau$.
    Furthermore, let $\pi \in \pos(t_i)$.
    \begin{itemize}
        \item[(a)] If $\pi < \tau$ or $\pi \bot \tau$ (i.e., $\pi$ is above or parallel to $\tau$),        
          then there is an edge from $(i,\pi)$ to $(i+1,\pi)$. The reason is that
          if position $\pi$ is annotated in $t_i$,
        then in a chain it would remain annotated in $t_{i+1}$.
      \item[(b)] For $\pi=\tau$,
        there are at most two outgoing edges from $(i,\pi)$ if
        the rule is in $\R^{=}$ and at most one edge if the rule is in $\R$. These edges
        lead to nodes of the
        form
        $(i+1,\pi.\alpha)$ for $\alpha \in \pos_{\Sigma}(r)$.
        The reason is that rules in $\ADPairMain{\R}$ contain at most one annotation and
        rules in $\ADPairBase{\R}$ contain at most two annotations in their right-hand sides. Moreover, 
        if $\pi$ is annotated in $t_{i}$, then we may create annotations
        at the positions $\pi.\alpha$ in the right-hand side 
        of the used ADP.
      \item[(c)] For every variable position
$\alpha_\ell \in \pos_\VSet(\ell)$, either there are no outgoing edges from any node of
        the form $(i,\tau.\alpha_\ell.\beta)$ with
 $\beta \in \NN^*$, or there is a 
        position $\alpha_r \in \pos_{\VSet}(r)$ with
        $r|_{\alpha_r} = \ell|_{\alpha_\ell}$ and for all $\beta \in \NN^*$ with
        $\alpha_\ell.\beta \in \pos(\ell\sigma)$, there is an edge from 
        $(i,\tau.\alpha_\ell.\beta)$ to $(i+1,\tau.\alpha_r.\beta)$. This captures the
        behavior of VRFs.
        \item[(d)] For all other positions $\pi \in \pos(t_i)$, there is no outgoing edge from the node $(i, \pi)$.
    \end{itemize}
\end{definition}

Due to \Cref{def:orig}, we have the following connection between origin graphs and
chains. For every origin graph for a
rewrite sequence $t_0 \to_{\R \cup \R^{=}} t_1 \to_{\R \cup \R^{=}} \ldots$ and every
$\tilde{t}_0$ such that $\flat(\tilde{t}_0) = t_0$, there is a chain
 $\tilde{t}_0 \tored{}{}{\ADPairMain{\R} \cup \ADPairBase{\R^{=}}} \tilde{t}_{1}
\tored{}{}{\ADPairMain{\R} \cup \ADPairBase{\R^{=}}} \ldots$ with
$\flat(\tilde{t}_i) = t_i$ for all $i \in \NN$ such that:
\[ \begin{array}{c}
\text{$\pi \in 
\posT(\tilde{t}_i)$}\\
\text{iff}\\
\text{there is a path in the origin graph
from $(0,\tau)$ to $(i,\pi)$ for some $\tau
\in \posT(\tilde{t}_0)$}
  \end{array}\]

\begin{example}\label{graph1}
    Let $\R \cup \R^{=}$ have 
    the rules $\ta(x) \to \tb(x)$, $\tg(x,x) \to \ta(x)$,   $\tb(x) \to \tb(x)$, and
    consider
    the rewrite sequence
    $\Theta:\tg(\underline{\ta(\O)}, \tb(\O)) \to_{\R\cup \R^{=}} \underline{\tg(\tb(\O), \tb(\O))} \to_{\R\cup \R^{=}} \ta(\tb(\O)) \to_{\R\cup \R^{=}} \ldots$. So here we
    have $t_0 = \tg(\ta(\O), \tb(\O))$, $t_1 = \tg(\tb(\O), \tb(\O))$, and $t_2 =
    \ta(\tb(\O))$.
    The following is a possible origin graph for $\Theta$.
    \[ \xymatrix @-1.5pc {
        t_0=&\tg \ar@{-}[d] &(& \ta \ar@{-}[d]&(&\O\ar@{-}[d]&), & \tb\ar@{-}[d] & (&\O\ar@{-}[d]&)) \\
        t_1=&\tg \ar@{-}[d] &(    & \tb \ar@{-}[d]&(&\O\ar@{-}[d]&), & \tb\ar@{-}[dllll] & (&\O\ar@{-}[dllll]&)) \\
        t_2=&\ta            &( & \tb           &(&\O          &)) &                   &  &  &
    }
    \]
\end{example}

We can now prove the chain criterion in the relative setting.
In the following,
we often use the notation $\pos_{f}(t)$ instead of $\pos_{\{f\}}(t)$
for a term $t$ and a single symbol or variable $f \in \Sigma \cup \VSet$.

\RelChainCriterion*

\begin{myproof}
    \underline{Completeness}:
    Assume that the ADP problem $(\ADPairMain{\R}, \ADPairBase{\R^{=}})$ is not SN.
    Then, there exists an infinite $(\ADPairMain{\R}, \ADPairBase{\R^{=}})$-chain,
    i.e., an infinite rewrite sequence $t_0 \tored{}{}{\ADPairMain{\R} \cup \ADPairBase{\R^{=}}} t_{1} \tored{}{}{\ADPairMain{\R} \cup \ADPairBase{\R^{=}}} \ldots$
    that uses an infinite number of rewrite steps with $\ADPairMain{\R}$ and Case $(\mathbf{pr})$.
    
    By removing all annotations we obtain the rewrite sequence
    $\flat(t_0) \to_{\R \cup \R^{=}} \flat(t_1) \to_{\R \cup \R^{=}} \ldots$
    that uses an infinite number of rewrite steps with $\R$.
    Hence, $\R / \R^{=}$ is not SN either.

    \smallskip

    \noindent
    \underline{Soundness}:
    Assume that $\R / \R^{=}$ is not SN.
    Then there exists an infinite sequence $\Theta: t_0 \to_{\R \cup \R^{=}} t_1 \to_{\R \cup \R^{=}} \ldots$ 
    that uses an infinite number of $\R$-rewrite steps.

    We will define a sequence $\tilde{t}_0, \tilde{t}_1, \ldots$ of annotated terms 
    such that $\flat(\tilde{t}_i) = t_i$ and
    $\tilde{t}_i \tored{}{}{\ADPairMain{\R} \cup \ADPairBase{\R^{=}}} \tilde{t}_{i+1}$ for all
    $i \in \IN$, 
    where we use an infinite number of rewrite steps with $\ADPairMain{\R}$ and Case $(\mathbf{pr})$.
    This is an infinite $(\ADPairMain{\R}, \ADPairBase{\R^{=}})$-chain, and hence, $(\ADPairMain{\R}, \ADPairBase{\R^{=}})$ is not SN either.

    W.l.o.g., let $t_0$ be a minimal term that is
    non-terminating w.r.t.\ $\R/\R^{=}$, i.e., there exists no proper subterm of $t_0$ that starts 
    a rewrite sequence which uses an infinite number of $\R$-steps.
    Such a minimal term exists, since there are only finitely many subterms of $t_0$.
    Next, we prove that there exists an origin graph for $\Theta$ with a path from
    $(0,\varepsilon)$ to some node $(i,\pi)$
    and a path  from  $(0,\varepsilon)$ to some node  $(i+1,\pi')$ with $i \in \IN$ such that
    the rewrite step $t_i \to_\R t_{i+1}$ takes place at position $\pi$, and
    $t_{i+1}|_{\pi'}$ is a minimal non-terminating term w.r.t.\ $\R/\R^{=}$.
    Then, according to the definition of the origin graph, there exists a chain
    $\tilde{t}_0, \ldots, \tilde{t}_i,
    \tilde{t}_{i+1}$ with $\flat(\tilde{t}_j) = t_j$ for all $1 \leq j \leq i+1$
    such that the positions $\pi$ in $\tilde{t}_i$ and $\pi'$ in $\tilde{t}_{i+1}$ are annotated.
    Hence, the  rewrite step
 $\tilde{t}_i \tored{}{}{\ADPairMain{\R} \cup \ADPairBase{\R^{=}}} \tilde{t}_{i+1}$ 
    is performed with $\ADPairMain{\R}$ and Case $(\mathbf{pr})$.
    Furthermore, there is a minimal non-terminating subterm of $\tilde{t}_{i+1}$ whose root is annotated.
    Thus, we can perform the whole construction again to create another chain
    starting in $\tilde{t}_{i+1}$
    that ends in a rewrite step with $\ADPairMain{\R}$ and Case $(\mathbf{pr})$.
    By repeating this construction infinitely often, we generate our desired infinite chain.

    It remains to prove that such an origin graph exists for
    every minimal non-terminating term $t_0$.
Let $k$ be the arity of 
$t_0$'s root symbol.
Then for $i \in \NN$, by induction we  now define
$t_i^1,\ldots,t_i^k$ such that $t_i^1,\ldots,t_i^k \triangleleft t_i$ at parallel
positions and such that $t_0^j \to_{\R \cup \R^=}^{\leq 1} t_1^j \to_{\R \cup \R^=}^{\leq
  1} \ldots$ for all $1 \leq j \leq k$, where $\to_{\R \cup \R^=}^{\leq 1} = (\to_{\R \cup \R^=} \cup =)$, i.e., 
$\to_{\R \cup \R^=}^{\leq 1}$ is the reflexive closure of $\to_{\R \cup \R^=}$. In addition, for $i \in \NN$ we define 
a non-empty context $C_i$
such that $t_i = C_i[q_{1,1},\ldots,q_{1,h_1}, \ldots, q_{k,1}, \ldots, q_{k, h_k}]$ for
subterms $q_{j,1},\ldots,q_{j,h_j} \trianglelefteq t_i^j$ at parallel positions for all $1
\leq j \leq k$.
Moreover, if $i > 0$, then for all pairs of positions $\pi_1, \pi_2 \in \pos_\Sigma(C_i)$
    we show that one
    can construct an origin graph with paths from $(0,\varepsilon)$ to $(i, \pi_1)$ and
from $(0,\varepsilon)$ to
    $(i, \pi_2)$.

  Note that there exists an $i \in \NN$ such that the rewrite
    step from $t_i$ to $t_{i+1}$ is done with an $\R$-rule at a position in $C_i$. 
    The reason is that otherwise,
infinitely many $\R$-steps would be applied on terms of the form $q_{j,b}$. But this would
mean that there exists a $t_0^j$ such that the sequence 
 $t_0^j \to_{\R \cup \R^=}^{\leq 1} t_1^j \to_{\R \cup \R^=}^{\leq
  1} \ldots$ has infinitely many $\R$-steps. However, this 
would be a contradiction to the minimality of $t_0$ because
 $t_0^j$ is a proper subterm of $t_0$.
So we define $t_i^1,\ldots,t_i^k$ and $C_i$ for all $i \geq 0$ until we reach the first $i$
where
an $\R$-step is performed with a redex at a position in $C_i$.

    We start with the case $i=0$. 
Let
    $t_0^{1}, \ldots, t_0^{k}$ be the subterms of $t_0$ at positions $1, \ldots, k$
    (i.e., $t_0^j = t_0|_j$ for all $1 \leq j \leq k$).
    Then, we have $t_0 = C_0[t_0^{1}, \ldots, t_0^{k}]$ for the context $C_0$ that only consists
    of $t_0$'s root symbol applied to $k$ holes.

    In the induction step, we have
    $t_i \to_{\R \cup \R^{=}} t_{i+1}$ using a position $\pi$, a substitution $\sigma$, and a
    rule $\ell \to r$ with $t_i|_{\pi} = \ell \sigma$ and $t_{i+1} = t_i[r
    \sigma]_{\pi}$. Here, we have two cases:
    \begin{itemize}
        \item If $\pi \notin \pos(C_i) \setminus \pos_{\Box}(C_i)$, 
          then $\pi$ must be in some $q_{j,b} \trianglelefteq t_{i}^{j}$.
          So there is an $\alpha \in \pos_{\Box}(C_i)$
such that $\pi =  \alpha.\beta$ for some  $\beta \in \IN^*$.
Hence, we have $t_i|_\pi = 
C_i[q_{1,1},\ldots,q_{1,h_1}, \ldots, q_{k,1}, \ldots, q_{k, h_k}]|_\pi = q_{j,b}|_\beta$
for some $q_{j,b} = t_i^j|_\gamma$. 
     Here, we simply perform the rewrite step on this term, and the context and the other terms remain the same.
        Hence, we have $C_{i+1} = C_{i}$, $t_{i+1}^{j} = t_i^{j}[r \sigma]_{\gamma.\beta}$ and
        $t_{i+1}^{j'} = t_i^{j'}$ for all $1 \leq j \leq k$ with $j' \neq j$.
        Using the subterms $q'_{1,1}, \ldots, q'_{1, h_1'}, \ldots,
        q'_{k,1}, \ldots, q'_{k, h_k'}$ with $q'_{j,b} = q_{j,b}[r \sigma]_{\beta}$ and
        $q'_{c,d} = q_{c,d}$ for all $1 \leq c \leq k$ and $1 \leq d \leq h_c$
        with $(c,d) \neq (j,b)$,
        we finally get $t_{i+1} = C_{i+1}[q'_{1,1}, \ldots, q'_{1, h_1'}, \ldots,
        q'_{k,1}, \ldots, q'_{k, h_k'}]$, as desired.
        
        It remains to prove that our claim on the paths in the origin graph is still satisfied.
        For all $\tau_1, \tau_2 \in \pos_\Sigma(C_{i+1}) = \pos_\Sigma(C_{i})$, by the induction
        hypothesis there
        exists an origin graph with
        paths from $(0,\varepsilon)$ to $(i, \tau_1)$ and from
$(0,\varepsilon)$ to $(i, \tau_2)$. Since the origin graph has  edges from
        $(i, \tau_1)$ to $(i+1,\tau_1)$ and from
        $(i, \tau_2)$ to $(i+1,\tau_2)$
        by \Cref{def:orig} since $\tau_1$ and $\tau_2$ are above or parallel to $\pi$, 
        there are also paths from $(0,\varepsilon)$ to $(i+1, \tau_1)$ and from
        $(0,\varepsilon)$ to $(i+1, \tau_2)$.

        \item Now we consider the case $\pi \in \pos(C_i) \setminus \pos_{\Box}(C_i)$.
        If the step $t_i \to_{\R \cup \R^{=}} t_{i+1}$ is an $\R$-step, then we stop, because
        we reached the first $\R$-step where the redex is at a position in $C_i$.

        So we now have $t_i \to_{ \R^{=}} t_{i+1}$. We define $t_{i+1}^j = t_i^j$ for all
        $1 \leq j \leq k$, but we still need to define the context and the subterms for
        each $t_{i+1}^j$. We will now define a suitable VRF
        $\varphi:\pos_{\VSet}(\ell)\to \pos_{\VSet}(r) \cup \{ \bot \}$ step by step, where we initialize
        $\varphi$ to yield $\bot$ for all arguments.
        For every $\rho \in \pos_{\VSet}(\ell)$,
        if possible,
        we let
        $\varphi(\rho) \in \pos_{\VSet}(r)$ be a position of $r$ that is not yet in the
        image of $\varphi$ and where
        $\ell|_{\rho} = r|_{\varphi(\rho)}$. If there is no such position
        of $r$, then we keep $\varphi(\rho) = \bot$.
 
        Let $\varphi|^{\pos_{\VSet}(r)}$
denote the restriction of $\varphi$ to the codomain $\pos_{\VSet}(r)$ (i.e.,
$\varphi|^{\pos_{\VSet}(r)}$
is only defined on those $\rho \in \pos_{\VSet}(\ell)$ where
$\varphi(\rho) \neq \bot$). Then 
 $\varphi|^{\pos_{\VSet}(r)}$ is surjective, since
        rules of $\R^=$ must not be duplicating, and injective, since we only extend the
        function $\varphi$ if a position of a variable in $r$ was not already in the image
        of $\varphi$.
        Let $\{\rho_1, \ldots, \rho_w\} \subseteq \pos_{\VSet}(\ell)$
        be those positions
        of variables from $\ell$  in the context $C_i$ that are no holes
        (i.e., where $\pi.\rho_z \in
        \pos(C_i)\setminus
        \pos_{\Box}(C_i)$)
        and where 
        $\varphi(\rho_z) \neq \bot$ for all $1 \leq z \leq w$.
        Then we define the new context $C_{i+1}$ as 
        $C_{i+1} = C_i[r \delta_{\Box}]_{\pi}[C_i|_{\pi.\rho_1}]_{\pi.\varphi(\rho_1)} \ldots [C_i|_{\pi.\rho_{w}}]_{\pi.\varphi(\rho_{w'})}$ 
        using the substitution $\delta_{\Box}$ that maps every variable to $\Box$.
        So $C_{i+1}$ results from $C_i$ by replacing the subterm at position $\pi$ by $r$
        where all variables are substituted with $\Box$, and then restoring
        the part of the context that was inside the substitution.

        Next, we need to define the subterms
        $q'_{j,1}, \ldots, q'_{j, h_j'}$ of $t_{i+1}^j$ 
        such that $t_{i+1} = C_{i+1}[q'_{1,1}, \ldots, q'_{1, h_1'}, \ldots,
          q'_{k,1}, \ldots, q'_{k, h_k'}]$.
        Let $\kappa$ be the position of some $q_{j,b}$ in $t_i$.
        First, consider the case that
        for some $\alpha_{\ell} \in \pos_{\VSet}(\ell)$ and $\beta \in \IN^*$ we have $\pi.\alpha_{\ell}.\beta = \kappa$, 
        i.e., $q_{j,b}$ is completely inside the substitution.
        If $\varphi(\alpha_{\ell}) = \bot$, then we remove the subterm $q_{j,b}$ in $t_{i+1}$.
        Otherwise, $q_{j,b}$ is one of the subterms $q'_{j,b'}$ and it moves from position $\pi.\alpha_{\ell}.\beta$
        in $t_i$ to position $\pi.\varphi(\alpha_{\ell}).\beta$ in $t_{i+1}$.
        Note that there are no two such $q_{j_1,b_1}$, $q_{j_2,b_2}$ that move to the same position,
        due to injectivity of $\varphi$.
        Second, $q_{j,b}$ is also one of the subterms $q'_{j,b'}$ if the
        position $\kappa$ is parallel to $\pi$.
        Here, the position of $q_{j,b}$ in $t_{i+1}$ remains the same as in $t_i$.
        Third, we consider the case that 
        there exist some $\alpha_{\ell} \in \pos_{\VSet}(\ell)$ such that $\kappa < \pi.\alpha_{\ell}$,
        i.e., $q_{j,b}$ is a subterm of the redex but not completely inside the substitution.
        For all such $\alpha_{\ell}$, let $\chi_{\alpha_\ell} \in \IN^*$ such that
        $\kappa.\chi_{\alpha_\ell} = \pi.\alpha_{\ell}$.
        Instead of the subterm $q_{j,b}$, we now 
        use the subterms $q_{j,b}|_{\chi_{\alpha_\ell}}$ for all those $\alpha_\ell$ with $\varphi(\alpha_{\ell}) \neq \bot$. 
        Now $q_{j,b}|_{\chi_{\alpha_\ell}}$ is a subterm of $t_{i+1}$ at position $\pi.\varphi(\alpha_{\ell})$.
        All in all, we get $t_{i+1} = C_{i+1}[q'_{1,1}, \ldots, q'_{1, h_1'}, \ldots,
        q'_{k,1}, \ldots, q'_{k, h_k'}]$ and $q'_{j,1},\ldots,q'_{j,h_j'} \trianglelefteq t_{i+1}^j$ at parallel positions for all $1
        \leq j \leq k$.

        Finally, we prove that our claim on the paths in the origin graph is
        still satisfied.
        Consider two positions $\tau_1, \tau_2 \in \pos_\Sigma(C_{i+1})$, and let $\tau \in \{\tau_1, \tau_2\}$.
        If $\tau$ is above or parallel to $\pi$, then by the induction
        hypothesis there exists an origin graph with
        a path from $(0,\varepsilon)$ to $(i, \tau)$. Since the origin graph has an edge from
        $(i, \tau)$ to $(i+1,\tau)$ by \Cref{def:orig}, 
        there is a path from $(0,\varepsilon)$ to $(i+1, \tau)$.
        If $\tau$ has the form $\pi.\alpha$ with $\alpha \in \pos_\Sigma(r)$ in $C_{i+1}$,
        then by the induction
        hypothesis there is an origin graph with
        a path from $(0,\varepsilon)$ to $(i, \pi)$. Since the origin graph can be chosen to
        have an edge from
        $(i, \pi)$ to $(i+1,\pi.\alpha)$ by \Cref{def:orig}, 
        there is a path from $(0,\varepsilon)$ to $(i+1,\pi.\alpha)$.
        Note that if both $\tau_1$ and $\tau_2$ have such a form (i.e., $\tau_1 =
        \pi.\alpha_1$ and $\tau_2 = \pi.\alpha_2$ with $\alpha_1, \alpha_2 \in
        \pos_\Sigma(r)$),
        then the origin graph can be chosen to have edges from $(i,\pi)$ to both 
        $(i+1,\pi.\alpha_1)$ and $(i+1,\pi.\alpha_2)$, as we can have  two such edges for
        a rewrite step with a relative rule from $\R^=$.
        Finally, if $\tau$ has the form $\pi.\alpha_r.\beta$ with
        $\alpha_r \in \pos_\VV(r)$ in $C_{i+1}$,
        then since the image of $\varphi$ consists of all variable positions from
         $\pos_\VV(r)$ (due to non-duplication of $\R^=$),
        there is a $\rho \in \pos_\VV(\ell)$ with
        $\ell|_{\rho} = r|_{\varphi(\rho)}$ where $\varphi(\rho)
        = \alpha_r$.
        By the induction hypothesis, there is an origin graph with
        a path from $(0,\varepsilon)$ to $(i, \pi.\rho.\beta)$ and by 
        \Cref{def:orig}, the origin graph can be chosen to have an edge from
        $(i, \pi.\rho.\beta)$ to $(i+1, \pi.\alpha_r.\beta)$.
        All in all, there exists an origin graph with paths from $(0,\varepsilon)$ to $(i, \tau_1)$
        and from $(0,\varepsilon)$ to $(i, \tau_2)$.
    \end{itemize}

So we have shown that 
    there exists an origin graph for $\Theta$ with a path from
    $(0,\varepsilon)$ to some node $(i,\pi)$
    such that
    the rewrite step $t_i \to_\R t_{i+1}$ takes place at position $\pi$ in the context
    $C_i$. Moreover, for any further position $\pi' \in \pos_\Sigma(C_i)$, in this origin
    graph there is also a path from   $(0,\varepsilon)$ to some node $(i,\pi')$.
In a similar way, one can also extend the construction one step further to define 
$t_{i+1}^1,\ldots,t_{i+1}^k$ and a non-empty context $C_{i+1}$ such that
 $t_{i+1} = C_{i+1}[q_{1,1},\ldots,q_{1,h_1}, \ldots, q_{k,1}, \ldots, q_{k, h_k}]$ for
subterms $q_{j,1},\ldots,q_{j,h_j} \trianglelefteq t_{i+1}^j$ (not necessarily at parallel positions, 
since the used $\R$-step may be duplicating).
%
Moreover,  for all  positions $\pi_1 \in \pos_\Sigma(C_i)$ and  $\pi_2 \in \pos_\Sigma(C_{i+1})$
   one
    can construct an origin graph with paths from $(0,\varepsilon)$ to $(i, \pi_1)$ and
from $(0,\varepsilon)$ to
    $(i+1, \pi_2)$.
For this last step,
 the cases of the proof are exactly the same, the only difference is that from node $(i,\pi)$,
    there may only be a single outgoing edge (since the used rule was in $\R$).
    But as we are only looking for a single position $\pi_2$ in $\pos_\Sigma(C_{i+1})$
    (and no pairs of positions anymore), 
    this suffices for the construction.

Then we have shown that the desired origin graph exists, by choosing $\pi_1$ to be $\pi$
(the position of the redex in the $\R$-step from $t_i$ to $t_{i+1}$)
and by choosing $\pi_2$ to be the position of a minimal non-terminating subterm of
$t_{i+1}$. Note that this position must be in $\pos_\Sigma(C_{i+1})$, because all subterms
$q_{j,b}$ are terminating w.r.t.\ $\R/\R^=$.    
\end{myproof}

For the following proof,
recall that $t \trianglelefteq_{\#}^{\tau} s$ if
$\tau \in \posT(s)$ and $t = \flat(s|_\tau)$.

\DerelProcOne*

\begin{myproof}
Completeness of any processor that yields the empty set is trivial. So we only have to
consider soundness.
  
    \underline{Only if}:
    Assume that $(\DP{\C{P}},\flat(\C{P} \cup \C{P}^{=}))$ is not SN.
    Then
there exists an infinite sequence $t_0, t_1, t_2, \ldots$ 
    with $t_i \rootto_{\DP{\C{P}}} \circ \to_{\flat(\C{P} \cup \C{P}^{=})}^* t_{i+1}$ for all $i \in \IN$.
    We now create a sequence $s_0, s_1, \ldots$ of annotated terms such that 
    $s_i \tored{}{(\mathbf{pr})}{\C{P}} \circ \tored{}{(\mathbf{r}) \, *}{\C{P} \cup
      \C{P}^{=}} s_{i+1}$
and $\flat(t_i) \trianglelefteq_{\#} s_i$    
    for all $i \in \IN$, which  by \Cref{theorem:relative-chain-crit} implies that $(\C{P},\C{P}^{=})$ is not SN, i.e., the
    processor is not sound.
    Initially, we start with the term $s_0 = t_0$.
    We have 
    \[
        t_0 \rootto_{\DP{\C{P}}} t_{0,1} \to_{\flat(\C{P} \cup \C{P}^{=})} t_{0,2} \to_{\flat(\C{P} \cup \C{P}^{=})} \ldots \to_{\flat(\C{P} \cup \C{P}^{=})} t_{1}
    \]
    In the first rewrite step, there is a DP $\ell^\# \to t^\# \in \DP{\C{P}}$ and a substitution $\sigma$
    such that $t_0 = \ell^\# \sigma$ and $t_{0,1} = t^\# \sigma$.
    This DP $\ell^\# \to t^\#$ results from some ADP $\ell \to r \in \C{P}$ with
    $t \trianglelefteq_{\#}^{\tau} r$
    for some position $\tau \in \pos(r)$.
    We can rewrite $s_0$ with $\ell \to r$ and the substitution $\sigma$ at the root
    resulting in $s_{0,1} = r\sigma$ with
    $\flat(t_{0,1}) \trianglelefteq_{\#}^{\tau}
    r\sigma = s_{0,1}$. 
    Then, we mirror each step that takes place at position $\pi$ in $t_{0,i}$ at position $\tau.\pi$ in $s_{0,i}$.
    To be precise, if we have $t_{0,i+1} = t_{0,i}[r' \sigma']_{\pi}$ using a rule $\ell' \to \flat(r') \in \flat(\C{P} \cup \C{P}^{=})$, the substitution $\sigma'$, and the position $\pi$, then we have $\flat(t_{0,i}) \trianglelefteq_{\#}^{\tau} s_{0,i}$, and can rewrite $s_{0,i}$
    with the ADP $\ell' \to r'$, the substitution $\sigma'$, and the position $\tau.\pi$.
    We get $\flat(s_{0,i+1}) = \flat(s_{0,i}[r' \sigma']_{\tau.\pi})$
    with $\flat(t_{0,i+1}) = \flat(t_{0,i}[r' \sigma']_{\pi}) \trianglelefteq_{\#}^{\tau}  s_{0,i+1}$.
    In the end, we have $\flat(t_1) \trianglelefteq_{\#}^{\tau} s_1$.

    We can now repeat this for each $i \in \IN$ and result in our desired chain.
    
    \smallskip

    \noindent
    \underline{If}:
    Assume that the processor is not sound and that $(\C{P},\C{P}^{=})$ is not SN.
    Then,  by \Cref{theorem:relative-chain-crit} there exists an infinite sequence $t_0, t_1, t_2, \ldots$ 
    of annotated terms such that 
    $t_i \tored{}{(\mathbf{pr})}{\C{P}} \circ \tored{}{*}{\C{P} \cup \C{P}^{=}} t_{i+1}$ 
    for all $i \in \IN$.
    W.l.o.g., let $t_0$ contain only a single annotation (we only need the annotation for the position that leads to infinitely many $\C{P}$-steps with Case $(\mathbf{pr})$).
    We now create a sequence $s_0, s_1, \ldots$ such that 
    $s_i \rootto_{\DP{\C{P}}} \circ \to_{\flat(\C{P} \cup \C{P}^{=})}^* s_{i+1}$
    and $\flat(s_i) \trianglelefteq_{\#} t_i$
    for all $i \in \IN$, 
    which implies that $(\DP{\C{P}},\flat(\C{P} \cup \C{P}^{=}))$ is not SN.
    Due to the condition $\flat(\C{P}^=) = \C{P}^=$, we have 
    \[  
        t_0 \tored{}{(\mathbf{pr})}{\C{P}} t_{0,1} \tored{}{(\mathbf{r})}{\C{P} \cup \C{P}^{=}}  t_{0,2} \tored{}{(\mathbf{r})}{\C{P} \cup \C{P}^{=}} \ldots \tored{}{(\mathbf{r})}{\C{P} \cup \C{P}^{=}} t_{1}
    \]
    The reason that we have no $\tored{}{(\mathbf{pr})}{\C{P}^{=}}$-steps is that
all terms $t_i$ and $t_{i,j}$ 
only contain a single annotation.
  Since $\flat(\C{P}^=) = \C{P}^=$, we only have annotations in $\C{P}$, where every
  right-hand side of a rule only has at most a single annotation.
  Thus, no term $t_i$ or $t_{i,j}$ can have more than one annotation.
    If we would rewrite at the position of this annotation with a rule without
    annotations (e.g., from $\C{P}^{=}$), then we would not have any annotation left 
    and there would be no future $\tored{}{(\mathbf{pr})}{\C{P}}$-step possible anymore.

    In the first rewrite step, there is an ADP $\ell \to r \in \C{P}$, a substitution $\sigma$, and a position $\pi$
    such that $t_0|_{\pi} = \ell^\# \sigma$ and $t_{0,1} = t_{0}[\anno_{\rho}(r \sigma)]_{\pi}$
    for some $\rho \in \pos(r)$.
    Initially, we start with the term $s_0 = \ell^\# \sigma$.
    We can rewrite $s_0$ with $\ell^\# \to t^\# \in \DP{\C{P}}$ for
    $t \trianglelefteq_{\#}^{\rho} r$ and the substitution $\sigma$
    resulting in $s_{0,1} = t^\# \sigma$
    with $\flat(s_{0,1}) = t\sigma  \trianglelefteq_{\#}^{\rho}
    \anno_{\rho}(r \sigma)$, i.e.,
    $\flat(s_{0,1}) \trianglelefteq_{\#}^{\kappa}
    t_{0}[\anno_{\rho}(r \sigma)]_{\pi}
     = t_{0,1}$ for $\kappa = \pi.\rho$.
    Then, each rewrite step in $t_{0,i}$
    is mirrored  in $s_{0,i}$.
    
    To be precise, let
    $\flat(t_{0,i+1}) = \flat(t_{0,i}[r' \sigma']_{\tau})$ using a rule $\ell' \to r' \in \flat(\C{P} \cup
    \C{P}^{=})$, the substitution $\sigma'$, and the position $\tau$.
    If $\tau$ is above or parallel to the position $\kappa$ of the
    subterm that corresponds to our
    classical chain, then we do nothing and set $s_{0,i} =
    s_{0,i+1}$.
    However, a rewrite step on or above $\kappa$ might change the position of this subterm,
    i.e., the position that we denoted with $\kappa$ may be modified in each such step. Note
    however, that this subterm cannot be erased by such a step because then we would remove
    the only annotated symbol and the chain would become finite.
    If $\tau$ is below $\kappa$ (i.e., we have $\tau = \kappa.\pi'$ for some position $\pi'$), then we have $\flat(s_{0,i}) \trianglelefteq_{\#}^{\kappa} t_{0,i}$, and can rewrite $s_{0,i}$
    with the rule $\ell' \to \flat(r')$, the substitution $\sigma'$, and the position $\pi'$.
    If $\kappa$ denotes the current (possibly changed) position of the corresponding subterm,
    then
    we get $\flat(s_{0,i+1}) = \flat(s_{0,i}[r' \sigma']_{\pi'}) \trianglelefteq_{\#}^{\kappa}
    t_{0,i+1}$.
    Finally, $\tau$ cannot be $\kappa$, since then this would not be a $(\mathbf{r})$-rewrite step but a $(\mathbf{pr})$-rewrite step.
    In the end, we obtain $\flat(s_1) \trianglelefteq_{\#} t_1$.

    We can now repeat this for each $i \in \IN$ and result in our desired chain.
\end{myproof}

\DerelProcTwo*

\begin{myproof}
    \underline{Soundness}:
    By \Cref{def:relative-rewrite-chain}, $(\C{P},
    \C{P}^{=})$ is not SN iff there exists an infinite $(\C{P}, \C{P}^{=})$-chain. Such a
    chain would also be an infinite $(\C{P}\cup \C{P}^{=}_a, \C{P}^{=}_b)$-chain.
    There is an infinite $(\C{P}\cup \C{P}^{=}_a, \C{P}^{=}_b)$-chain iff there is an infinite
    $(\C{P}\cup \mathtt{split}(\C{P}^{=}_a), \C{P}^{=}_b)$-chain as we only need at most a single annotation in the main ADPs. 
    By \Cref{def:relative-rewrite-chain}, this is
    equivalent to non-termination of $(\C{P}\cup \mathtt{split}(\C{P}^{=}_a), \C{P}^{=}_b)$.
\end{myproof}

Before we prove the soundness and completeness of the dependency graph processor, we
present
another definition regarding origin graphs.
In \Cref{def:orig} we 
have seen that for a \emph{non-annotated} rewrite sequence $t_0 \to_{\R \cup \R^{=}} \ldots$ one
can obtain
several different origin graphs.
Now, we define the \emph{canonical} origin graph for an \emph{annotated} rewrite sequence
$t_0 \tored{}{}{\C{P} \cup \C{P}^{=}} \ldots$\
This graph 
represents the flow of the annotations in this sequence.

\begin{definition}[Canonical Origin Graph]\label{def:can_orig}
    Let $(\C{P}, \C{P}^=)$ be an ADP problem
    and let $\Theta: t_0 \tored{}{}{\C{P} \cup \C{P}^{=}} t_1 \tored{}{}{\C{P} \cup
      \C{P}^{=}} \ldots$\
    The \defemph{canonical origin graph} for $\Theta$
    has the nodes $(i,\pi)$  for all $i \in \IN$ and all $\pi \in \pos(t_i)$,
    and its edges are defined as follows:
  For $i \in \IN$,
    let
    the step  $t_i \tored{}{}{\C{P} \cup \C{P}^{=}} t_{i+1}$
 be performed using the rule
   $\ell \to r \in \C{P} \cup \C{P}^{=}$, the position $\tau$, the substitution $\sigma$, and the VRF $\varphi$.
    Furthermore, let $\pi \in \pos(t_i)$.
    \begin{itemize}
        \item[(a)] If $\pi < \tau$ or $\pi \bot \tau$ (i.e., $\pi$ is above or parallel to $\tau$),        
        then there is an edge from $(i,\pi)$ to $(i+1,\pi)$.
        \item[(b)] For $\pi=\tau$,
        there is an edge from $(i,\pi)$ to $(i+1,\pi.\alpha)$ for all $\alpha \in \posT(r)$.
        \item[(c)] If $\pi=\tau.\alpha_\ell.\beta$
        for a variable position $\alpha_\ell \in \pos_\VSet(\ell)$ and $\beta \in \NN^*$, then
        there is an edge from $(i,\tau.\alpha_\ell.\beta)$ to 
        $(i+1,\tau.\varphi(\alpha_\ell).\beta)$ if $\varphi(\alpha_\ell) \neq \bot$.
        \item[(d)] For all other positions $\pi \in \pos(t_i)$, there is no outgoing edge from the node $(i, \pi)$.
    \end{itemize}
    Moreover, if an ADP is applied with Case $(\mathbf{pr})$ at position $\tau$ in
   $t_i$ in   $\Theta$, then all edges originating in $(i, \tau)$ are labeled with this ADP. All other
    edges are not labeled.
\end{definition}

    So for $\R_2$ from \Cref{example:redex-creating}
    where $\ADPairMain{\R_2}$ consists of
    \begin{equation}\label{mainR2} 
        \ta \to \tb
    \end{equation}
    and $\ADPairBase{\R_2^{=}}$ consists of
    \begin{equation}\label{baseR2}
        \tf \to \tc(\tF,\tA),
    \end{equation}
    the chain from \Cref{ex:ADPs-for-redex-creation-1} yields the following canonical origin graph:
    \[ \xymatrix @-1.5pc @C=1pc{
    t_0=&   & &   & &\tF\phantom{,}\ar@{-}[dllll]_{\eqref{baseR2}}\ar@{-}[d]^{\eqref{baseR2}}\ar@{-}[drr]^{\eqref{baseR2}}     & &  \\
    t_1=&\td\ar@{-}[d]&(&   & &\tF,\ar@{-}[d] &    &\tA\ar@{-}[d]^{\eqref{mainR2}})\\
    t_2= &\td\ar@{-}[d]&(&   & &\tF\ar@{-}[dll]_{\eqref{baseR2}}\ar@{-}[d]_{\eqref{baseR2}}\ar@{-}[dr]^{\eqref{baseR2}},    & &\tb\ar@{-}[d])\\
    t_3=&\td\ar@{-}[d]&(&\td\ar@{-}[d]&(&\tF,\ar@{-}[d],&\tA\ar@{-}[d]^{\eqref{mainR2}})&\tb\ar@{-}[d])\\
    t_4=&\td&(&\td&(&\tF,&\tb),&\tb)
    }
    \]

    For $\R_3$ from \Cref{example:redex-creatingAbove}
    where $\ADPairMain{\R_3}$ consists of
    \begin{equation}\label{mainR3} 
        \ta(x) \to \tb(x)
    \end{equation}
    and $\ADPairBase{\R_2^{=}}$ consists of
    \begin{equation}\label{baseR3}
        \tf \to \tA(\tF),
    \end{equation}
    the chain from \Cref{ex:ADPs-for-redex-creation-2} yields the following canonical origin graph:
    \[ \xymatrix @-1.5pc @C=1pc{
    t_0= &  &  &  & & \tF\ar@{-}[dllll]_{\eqref{baseR3}}\ar@{-}[d]^{\eqref{baseR3}} &  \\
    t_1= & \tA\ar@{-}[d]_{\eqref{mainR3}} & ( &  & & \tF\ar@{-}[d] & ) \\
    t_2= & \tb\ar@{-}[d] & ( &  & & \tF\ar@{-}[dll]_{\eqref{baseR3}}\ar@{-}[d]^{\eqref{baseR3}} & ) \\
    t_3= & \tb\ar@{-}[d] & ( & \tA\ar@{-}[d]_{\eqref{mainR3}} & ( & \tF\ar@{-}[d] & )) \\
    t_4= & \tb & ( & \tb & ( & \tF & ))
    }
    \]

\RelativeDepGraphProc*

\begin{myproof}
    \underline{Completeness}:
    For every $(\C{P}', {\C{P}^{=}}')$-chain with $(\C{P}', {\C{P}^{=}}') \in
    \Proc_{\mathtt{DG}}(\C{P},\C{P}^{=})$ there exists a $(\C{P}, \C{P}^{=})$-chain with
    the same terms and possibly more annotations.
    Hence, if some ADP problem in $\Proc_{\mathtt{DG}}(\C{P},\C{P}^{=})$ is not SN, then
    $(\C{P}, \C{P}^{=})$ is
    not SN either.

    \smallskip

    \noindent
    \underline{Soundness}:
    By the definition of $\tored{}{}{}$, whenever there is a path from an edge labeled with
    an ADP $\ell \to r$ to an edge labeled with an ADP 
    $\ell' \to r'$ in the canonical origin graph, then there is a path from 
    $\ell \to r$ to
    $\ell' \to r'$ in the dependency graph. Since there only exist finitely many ADPs and the
    chain uses ADPs from $\PP$ with Case $(\mathbf{pr})$ infinitely many times, there are
    two cases:

    Either there exists a path in the canonical origin graph where infinitely many edges are labeled
    with an ADP from $\PP$. Then after finitely many steps, this path only uses edges
    labeled with ADPs from an SCC $\QQ$ of the dependency graph that contains an ADP from $\PP$
    (i.e., from a $\QQ
    \in \mathtt{SCC}_{\C{P}}^{(\C{P}, \C{P}^{=})}$).

    Otherwise, there is no such path in the canonical origin graph, but then there is a path in the
    canonical origin graph where infinitely many edges are labeled with ADPs from $\PP^=$ and this path
    generates infinitely many paths that lead to an edge labeled with an ADP from $\PP$. 
    Then after finitely many steps,
    this path only uses edges
    labeled with ADPs from a minimal lasso of the dependency graph (i.e., from a
    $\QQ \in \mathtt{Lasso}$).

    So in both cases, there exists a $\QQ \in \mathtt{SCC}_{\C{P}}^{(\C{P}, \C{P}^{=})} \cup \mathtt{Lasso}$
    such that the infinite path gives rise to an infinite $(\,(\C{P} \cap \QQ) \cup \flat(\C{P} \setminus \QQ), (\C{P}^= \cap \QQ) \cup \flat(\C{P}^{=} \setminus \QQ \,)\,)$-chain.
    Since the ADPs $\flat(\PP \setminus \QQ)$ are never used for steps with 
    Case $(\mathbf{pr})$ in this infinite chain, they can also be moved to the base ADPs. Thus, this is also an
    infinite $(\,\C{P} \cap \QQ, (\C{P}^= \cap \QQ)\cup \flat( \,(\C{P} \cup \C{P}^{=}) \setminus \QQ \,)\,)$-chain, 
    i.e., $(\,\C{P} \cap \QQ, (\C{P}^= \cap \QQ)\cup \flat( \,(\C{P} \cup \C{P}^{=}) \setminus \QQ \,)\,)$ is not SN either.
\end{myproof}

\RelRPP*

\begin{myproof}
    \underline{Completeness}:
    For every $(\C{P} \setminus \PP_{\succ}, (\C{P}^{=} \setminus \PP_{\succ}) \cup
        \flat(\PP_{\succ}))$-chain there exists a $(\C{P}, \C{P}^{=})$-chain with the same
        terms and possibly
        more annotations.
        Hence, if
$(\C{P} \setminus \PP_{\succ}, (\C{P}^{=} \setminus \PP_{\succ}) \cup
        \flat(\PP_{\succ}))$
 is not SN, then $(\C{P}, \C{P}^{=})$ is not SN either.

    \smallskip

    \noindent
    \underline{Soundness}:
    We start by showing that the conditions of the theorem extend to rewrite steps instead of just ADPs:
    \begin{enumerate}
        \item[(a)] If $s,t \in \TSet{\Sigma^\#}{\VSet}$ with $s \tored{}{}{\C{P} \cup \C{P}^{=}} t$, then $\subA(s) \succsim \subA(t)$.
        \item[(b)] If $s,t \in \TSet{\Sigma^\#}{\VSet}$ with $s \tored{}{}{\PP_{\succ}} t$ using Case $(\mathbf{pr})$, then $\subA(s) \succ \subA(t)$.
    \end{enumerate}
    For this, we extend $\subA(t)$ to terms with possibly more than two annotations by defining
    $\subA(t) = \Com{2}(r_1^\#,\ldots,\Com{2}(r_{n-1}^\#,r_n^\#)\ldots)$ 
    if $r_i \trianglelefteq_{\#}^{\pi_i} r$ for $n \geq 2$ and the
    positions $\pi_1, \ldots, \pi_n$ with $\pi_i <_{lex} \pi_{i+1}$ for all $1 \leq i < n$.

    \begin{itemize}
        \item[(a)] Assume that we have $s \tored{}{}{\C{P} \cup \C{P}^{=}} t$ using the
        ADP $\ell \to r$, the VRF $\varphi$, the position $\pi$, and the substitution
        $\sigma$. So we have $\flat(s|_{\pi}) = \ell \sigma$.
        Furthermore, let $\subA(s)$ contain the terms $s_1^\#, \ldots, s_n^\#$
        with annotated root symbols. If $n=0$, then we have $\subA(s) = \Com{0} = \subA(t)$ which
        proves the claim.

        Otherwise, 
        we partition $s_1,\ldots,s_n$ into several disjoint groups:

        Let $s_{i_1}, \ldots, s_{i_{n_1}}$ be all those $s_{i}$ that are at
        positions above $\pi$ in $s$ (i.e., these are those $s_i$ where  $\pi_{i} < \pi$).
        Let $s_{i_{n_1 + 1}}, \ldots, s_{i_{n_2}}$ be all those $s_{i}$
        that are at positions parallel to $\pi$ or on or below a variable position of $\ell$ that
        is ``considered'' by the VRF (i.e., those $s_i$ where
        $\pi_{i} \bot \pi$ or $\pi.\alpha \leq \pi_{i}$ for some $\alpha \in \pos_{\VSet}(\ell)$ with $\varphi(\alpha) \neq \bot$).
        Finally, let $s_{i_{n_2 + 1}}, \ldots, s_{i_{n_3}}$ be all those $s_{i}$
        that are below position $\pi$, but not on or below a variable position of $\ell$ that
        is ``considered'' by the VRF (i.e., those $s_i$ where
        $\pi_i = \pi.\alpha$ for some $\alpha \in \pos_{\Sigma}(\ell)$ with $\alpha \neq \epsilon$
        or $\pi.\alpha \leq \pi_{i}$ for some $\alpha \in \pos_{\VSet}(\ell)$ with $\varphi(\alpha) = \bot$).

        If the rewrite step takes place at a position that is not annotated (i.e., the symbol at
        position $\pi$ in $s$ is not annotated), then we have $n_3 =
        n$ and $\{ i_1, \ldots, i_{n_3}\} = \{1,\ldots, n\}$. 
        Otherwise, we have $n_3 = n-1$ and $\{s_{i_1},
        \ldots, s_{i_{n_3}}, \ell^\#\sigma\} = \{s_1, \ldots, s_n \}$.

        After the rewrite step with $\tored{}{}{\C{P} \cup \C{P}^{=}}$,
        $\subA(t)$
        contains the following terms with annotated root symbols:
        The terms
        $s_{i_{n_1 + 1}}, \ldots, s_{i_{n_2}}$ are unchanged and still contained in 
        $\subA(t)$. The terms $s_{i_{n_2 + 1}}, \ldots, s_{i_{n_3}}$ are removed, i.e., they are
        no longer in $\subA(t)$. 
        The terms $s_{i_1}, \ldots, s_{i_{n_1}}$ are replaced by 
        $s_{i_1}[\flat(r) \sigma]_{\tau_{1}}, \ldots, s_{i_{n_1}}[\flat(r) \sigma]_{\tau_{n_1}}$
        for appropriate positions $\tau_i \neq \varepsilon$.
        Furthermore, if
        the symbol at position $\pi$ in $s$ was annotated, then in addition, $\subA(t)$ contains
        the terms $r_1^\#\sigma, \ldots, r_m^\#\sigma$ where $r_j$ for $1 \leq j
        \leq m$ are all terms with $r_j \trianglelefteq_{\#} r$.
        Hence, if the symbol at position $\pi$ in $s$ was annotated, then there exist contexts $C, C', C''$
        containing no function symbol except $\Com{2}$ such that

\vspace*{-.2cm}
        
        {\small\[
            \begin{array}{lcl}
              \subA(s) & = & C[s_{i_1}, \ldots, s_{i_{n_1}},s_{i_{n_1 + 1}}, \ldots,
                s_{i_{n_2}},s_{i_{n_2 + 1}}, \ldots, s_{i_{n_3}},\ell^\#\sigma]\\
              & \succsim & C'[s_{i_1}, \ldots, s_{i_{n_1}},s_{i_{n_1 + 1}}, \ldots,
                s_{i_{n_2}},s_{i_{n_2 + 1}}, \ldots, s_{i_{n_3}},r_1^\#\sigma, \ldots,
                r_m^\#\sigma]\\
              &&\qquad\qquad \text{by $\ell^\# \succsim \subA(r)$, $\Com{}$-invariance, and $\Com{}$-monotonicity}\\
              & \succsim & C''[s_{i_1}, \ldots, s_{i_{n_1}},s_{i_{n_1 + 1}}, \ldots,
                s_{i_{n_2}},r_1^\#\sigma, \ldots,
                r_m^\#\sigma]\\
               & \succsim & C''[s_{i_1}[\flat(r) \sigma]_{\tau_{1}}, \ldots, s_{i_{n_1}}[\flat(r) \sigma]_{\tau_{n_1}},s_{i_{n_1 + 1}}, \ldots,
                s_{i_{n_2}},r_1^\#\sigma, \ldots,
                r_m^\#\sigma]\\
              &&\qquad\qquad \text{by $\ell \succsim \flat(r)$,
                $\Com{}$-invariance, and $\Com{}$-monotonicity}\\
                            & = & \subA(t)\!    
            \end{array}
            \]}

\vspace*{-.2cm}

        \noindent
        If the symbol at position $\pi$ in $s$ was not annotated, then we obtain

\vspace*{-.2cm}

        {\small\[
          \begin{array}{lcl}
    \subA(s) & = & C[s_{i_1}, \ldots, s_{i_{n_1}},s_{i_{n_1 + 1}}, \ldots,
                s_{i_{n_2}},s_{i_{n_2 + 1}}, \ldots, s_{i_{n_3}}]\\
                & \succsim & C'[s_{i_1}, \ldots, s_{i_{n_1}},s_{i_{n_1 + 1}}, \ldots,
                s_{i_{n_2}}]\\
               & \succsim & C'[s_{i_1}[\flat(r) \sigma]_{\tau_{1}}, \ldots, s_{i_{n_1}}[\flat(r) \sigma]_{\tau_{n_1}},s_{i_{n_1 + 1}}, \ldots,
                s_{i_{n_2}}]\\
              &&\qquad\qquad \text{by $\ell \succsim \flat(r)$, $\Com{}$-invariance, and $\Com{}$-monotonicity}\\
                            & = & \subA(t)\!    
            \end{array}
          \]}

        \item[(b)] Assume that we have $s \tored{}{}{\PP_{\succ}} t$ using the ADP $\ell \to r$, the VRF $\varphi$, the position $\pi$, the substitution $\sigma$, and we rewrite at an annotated position.
          So here, $\subA(s) \neq \Com{0}$.   Using the same notations as in (a), we get

\vspace*{-.2cm}

          {\small\[
            \begin{array}{lcl}
              \subA(s) & = & C[s_{i_1}, \ldots, s_{i_{n_1}},s_{i_{n_1 + 1}}, \ldots,
                s_{i_{n_2}},s_{i_{n_2 + 1}}, \ldots, s_{i_{n_3}},\ell^\#\sigma]\\
              & \succ & C'[s_{i_1}, \ldots, s_{i_{n_1}},s_{i_{n_1 + 1}}, \ldots,
                s_{i_{n_2}},s_{i_{n_2 + 1}}, \ldots, s_{i_{n_3}},r_1^\#\sigma, \ldots,
                r_m^\#\sigma]\\
              &&\qquad\qquad \text{by $\ell^\# \succ \subA(r)$, $\Com{}$-invariance, and $\Com{}$-monotonicity}\\
              & \succsim & C''[s_{i_1}, \ldots, s_{i_{n_1}},s_{i_{n_1 + 1}}, \ldots,
                s_{i_{n_2}},r_1^\#\sigma, \ldots,
                r_m^\#\sigma]\\
               & \succsim & C''[s_{i_1}[\flat(r) \sigma]_{\tau_{1}}, \ldots, s_{i_{n_1}}[\flat(r) \sigma]_{\tau_{n_1}},s_{i_{n_1 + 1}}, \ldots,
                s_{i_{n_2}},r_1^\#\sigma, \ldots,
                r_m^\#\sigma]\\
              &&\qquad\qquad \text{by $\ell \succsim \flat(r)$, $\Com{}$-invariance, and
                $\Com{}$-monotonicity}\\
                            & = & \subA(t)\!    
            \end{array}
        \]}
    \end{itemize}

    We can now prove soundness.
    Assume that $(\C{P}, \C{P}^{=})$ is not SN.
    Then there exists an infinite $(\C{P}, \C{P}^{=})$-chain $t_0 \tored{}{}{\C{P} \cup \C{P}^{=}} t_1 \tored{}{}{\C{P} \cup \C{P}^{=}} t_2 \ldots$
    If the chain uses an infinite number of rewrite steps with rules from $\PP_{\succ}$ and Case $(\mathbf{pr})$, 
    then $\subA(t_0) \succsim \subA(t_1) \succsim \subA(t_2) \succsim \ldots$
    would contain an infinite number of steps where the
    strict relation $\succ$ holds, which is a contradiction to well-foundedness of $\succ$, 
    as $\succ$ is compatible with $\succsim$.

    Hence, the chain only contains a finite number of $\tored{}{}{\PP_{\succ}}$-steps with
    Case $(\mathbf{pr})$. So there is an infinite suffix of the chain where only
    ADPs from $(\C{P} \setminus \PP_{\succ}) \cup (\C{P}^{=} \setminus \PP_{\succ})$ are
    used with
    Case $(\mathbf{pr})$.
    This means that $t_i \tored{}{}{\C{P} \cup \C{P}^{=}} t_{i+1} \tored{}{}{\C{P} \cup
      \C{P}^{=}} t_{i+2} \ldots$ is an infinite $((\C{P} \setminus \PP_{\succ}) \cup
        \flat(\PP \cap \PP_{\succ}), (\C{P}^{=} \setminus \PP_{\succ}) \cup
        \flat(\C{P}^{=} \cap \PP_{\succ}))$-chain, as ADPs that are only used for steps
        with Case $(\mathbf{r})$ do not need annotations.
Since the ADPs  $\flat(\PP \cap \PP_{\succ})$ are never used for steps with 
Case $(\mathbf{pr})$, they can also be moved to the base ADPs. Thus, this is also an
infinite $(\C{P} \setminus \PP_{\succ}, (\C{P}^{=} \setminus \PP_{\succ}) \cup
        \flat(\PP_{\succ}))$-chain, i.e., 
$(\C{P} \setminus \PP_{\succ}, (\C{P}^{=} \setminus \PP_{\succ}) \cup
        \flat(\PP_{\succ}))$ is not SN either.
\end{myproof}

Since ADPs are ordinary rewrite rules with annotations, we can also use
ordinary reduction orderings (that are closed under contexts)
to remove rules from an ADP problem completely.

\begin{restatable}[Rule Removal Processor]{theorem}{RelRRP}\label{thm:RelRRP}
    Let $(\C{P},\C{P}^{=})$ be an ADP problem, and let
    $(\succsim, \succ)$ be a
 reduction pair where $\succ$ is closed under contexts
    such that $\flat(\C{P} \cup \C{P}^{=}) \subseteq {\succsim}$.
  Moreover, let $\PP_{\succ} \subseteq \C{P} \cup \C{P}^{=}$ 
  such that $\flat(\C{P}_{\succ}) \subseteq {\succ}$.
Then
    $\Proc_{\mathtt{RR}}(\C{P},\C{P}^{=}) = \{(\C{P} \setminus \PP_{\succ}, \C{P}^{=} \setminus \PP_{\succ})\}$
 is sound and complete.
\end{restatable}

\begin{myproof}
    \underline{Completeness}:
    Every $(\C{P} \setminus \PP_{\succ}, \C{P}^{=} \setminus \PP_{\succ})$-chain 
    is also a $(\C{P},\C{P}^{=})$-chain.
    Hence, if $(\C{P} \setminus \PP_{\succ}, \C{P}^{=} \setminus \PP_{\succ})$ is not SN, then
    $(\C{P},\C{P}^{=})$ is not SN either.

    \smallskip

    \noindent
    \underline{Soundness}:
    Assume that $(\C{P},\C{P}^{=})$ is not SN.
    Let $t_0, t_1, \ldots$ be an infinite $(\C{P},\C{P}^{=})$-chain.
    Then, $t_i \tored{}{}{\C{P} \cup \C{P}^{=}} t_{i+1}$ for all $i \in \IN$.
    Since $\flat(\C{P} \cup \C{P}^{=}) \subseteq  {\succsim}$ and
    $\succsim$ is closed under contexts and substitutions, we obtain
     $\flat(t_0) \succsim \flat(t_1) \succsim \ldots$
    Assume for a contradiction that we use infinitely many steps with
$\tored{}{}{\C{P}_{\succ}}$.
    Then, $\flat(t_0) \succsim \flat(t_1) \succsim \ldots$
 would contain an infinite number of steps where the
 strict relation $\succ$ holds, because 
$\flat(\C{P}_{\succ}) \subseteq {\succ}$ and $\succ$  is also closed under contexts and
 substitutions. This
is a contradiction to well-foundedness of $\succ$, 
    as $\succ$ is compatible with $\succsim$.

  Hence, the chain only contains a finite number of $\tored{}{}{\PP_{\succ}}$-steps. So there is an infinite suffix of the chain where only
  ADPs from
$(\C{P} \cup \C{P}^{=})\setminus \PP_{\succ}$ are
  used.
      This means that $t_i \tored{}{}{\C{P} \cup \C{P}^{=}} t_{i+1} \tored{}{}{\C{P} \cup
        \C{P}^{=}} t_{i+2} \ldots$ is an infinite
      $(\C{P} \setminus \PP_{\succ}, \C{P}^{=} \setminus \PP_{\succ})$-chain.
Hence,
$(\C{P} \setminus \PP_{\succ}, \C{P}^{=} \setminus \PP_{\succ})$ is not SN either.
\end{myproof}